\documentclass[aps,pra,amssymb,superscriptaddress,twocolumn]{revtex4-1}
\usepackage[dvipdfmx]{graphicx}
\usepackage{bm}
\usepackage{amsmath}
\usepackage{ascmac}
\usepackage{color}
\usepackage{setspace}
\usepackage{amssymb}
\usepackage{latexsym}
\usepackage{mathtools}
\usepackage{lipsum}
\usepackage[colorlinks,linkcolor=blue,citecolor=blue]{hyperref}

\begin{document}

\title{Family-Vicsek Scaling of Roughness Growth in a Strongly Interacting Bose Gas} 

\author{Kazuya Fujimoto}
\affiliation{Institute for Advanced Research, Nagoya University, Nagoya 464-8601, Japan}
\affiliation{Department of Applied Physics, Nagoya University, Nagoya 464-8603, Japan}

\author{Ryusuke Hamazaki}
\affiliation{Department of Physics, University of Tokyo, 7-3-1 Hongo, Bunkyo-ku, Tokyo 113-0033, Japan}
\affiliation{Nonequilibrium Quantum Statistical Mechanics RIKEN Hakubi Research Team, RIKEN Cluster for Pioneering Research (CPR), RIKEN iTHEMS, Wako, Saitama 351-0198, Japan}

\author{Yuki Kawaguchi}
\affiliation{Department of Applied Physics, Nagoya University, Nagoya 464-8603, Japan}

\begin{abstract}
Family-Vicsek scaling is one of the most essential scale-invariant laws emerging in surface-roughness growth of classical systems. In this Letter, we theoretically elucidate the emergence of the Family-Vicsek scaling even in a strongly interacting quantum bosonic system by introducing a surface-height operator. This operator is comprised of a summation of local particle-number operators at a simultaneous time, and thus the observation of the surface roughness in the quantum many-body system and its scaling behavior are accessible to current experiments of ultracold atoms. 
\end{abstract}

\date{\today}

\maketitle

{\it Introduction.-}
Dynamic scaling is a hallmark of spatio-temporal scale-invariance, which plays a pivotal role in uncovering universal aspects behind complicated non-equilibrium phenomena \cite{cardy,hohenberg}. The typical examples are critical and coarsening dynamics \cite{bray2002,onuki2002,tauber2014}, in which essential information such as dimension and symmetry of a model classifies universality of the dynamics. Such universal dynamics has been widely observed in both classical \cite{C1,C2,C3,C4} and quantum systems \cite{KZM1,KZM2,KZM3,KZM4,dynamical_phase_transition1,NTFP_exp1,NTFP_exp2,Eigen2018}, being an arena for foundations of non-equilibrium statistical mechanics. 

\begin{figure}[b]
\begin{center}
\includegraphics[keepaspectratio, width=8.8cm,clip]{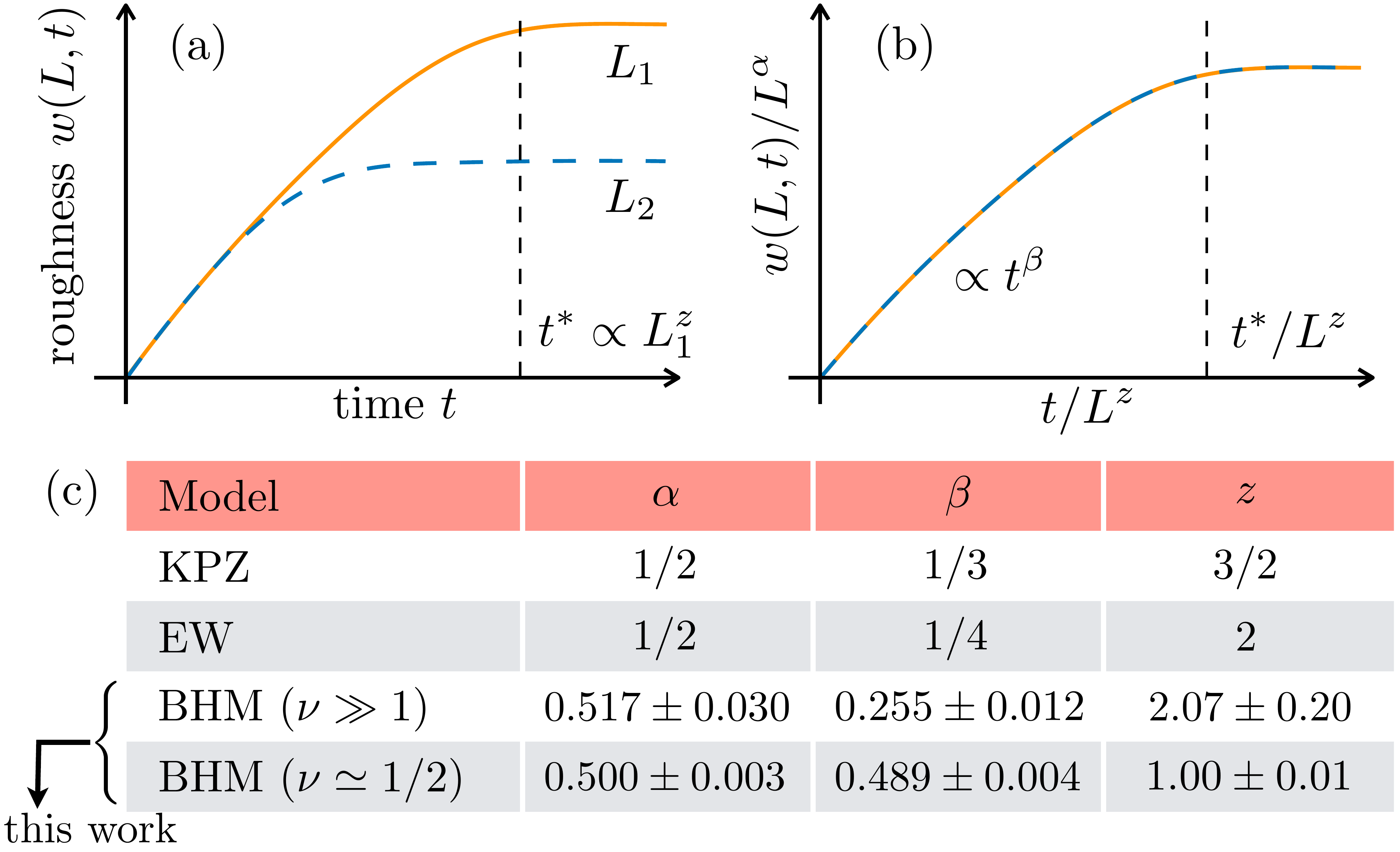}
\caption{Family-Vicsek scaling and its exponents. (a) Time-evolution of the surface roughness $w(L,t)$ for two different system sizes ($L_1>L_2$). The surface roughness grows in time and is finally saturated after the saturation time $t^*$. (b) Dynamic scaling of $w(L,t)$. When we normalize the ordinate and the abscissa by $L^{\alpha}$ and $L^z$, all curves collapse to a single one. The saturation time $t^{*}$ is scaled as $L^{z}$ with $z=\alpha / \beta$. (c) Exponents of Kardar-Parisi-Zhang (KPZ),  Edwards-Wilkinson (EW), and Bose-Hubbard models (BHM) ($\nu \gg1$ and $\nu \simeq 1/2$) with the filling factor $\nu$. This work finds the exponents of the BHM under the strong repulsive interaction.
\label{fig1} }
\end{center}
\end{figure}

Stochastic surface-growth is one of the long-standing universal dynamics discussed in classical non-equilibrium phenomena, where the roughness of the growing surface shows universal spatio-temporal scale-invariance \cite{barabasi1995fractal}. Consider a surface height $h(x,t)$ in a one-dimensional (1D) system with the linear size $L$. Then, the roughness $w(L,t)$ is quantified as the standard deviation of $h(x,t)$ from its spatial average. For a wide variety of stochastic processes, the roughness obeys a dynamic scaling law called the Family-Vicsek (FV) scaling \cite{Vicsek1984,Vicsek1985} (see Figs.~\ref{fig1} (a) and (b)):
\begin{eqnarray}
w(L,t) = s^{-\alpha} w(sL,s^{z}t) &\propto& 
 \begin{cases}
    t^{\beta} & ( t \ll t^*); \\
    L^{\alpha} & ( t^* \ll t)
  \end{cases}
\end{eqnarray}
with a constant $s$. Here, $t^*$ is a saturation time proportional to $L^{z}$, and $\alpha$, $\beta$, and $z=\alpha/\beta$ are power exponents featuring the universality of a stochastic surface growth model. The typical models are the Kardar-Parisi-Zhang (KPZ) model \cite{OKPZ} and the Edwards-Wilkinson (EW) model \cite{OEW}, whose universal exponents are shown in Fig.~\ref{fig1}(c). This kind of universality in stochastic surface growth has been extensively explored in various classical systems in the community of not only physics \cite{Prahofer2000,Prahofer2004,sasamoto2010,Calabrese2011,Takeuchi2010,Takeuchi2011,Takeuchi2012} but also mathematics \cite{Hairer} and biology \cite{Wakita1997,Hallatschek2007}.

It is then natural to ask whether or not such universal fluctuation dynamics appears in a relaxation process triggered by a parameter quench in quantum systems. Recent theoretical works study the KPZ universality class in quantum magnets by using the spin spatio-temporal correlation function \cite{Prosen2019,Moore2019,Ferdinand2019}. The similar scaling behavior emerges for the spatio-temporal correlation function of the density and phase fluctuations in a Bose-Einstein condensate, calculated by means of the (stochastic) Gross-Pitaevskii equation \cite{Kulkarni2013,Kulkarni2015,Mendl_2015,Altman2015,Liang2015,Zamora2017,Squizzato2018,Comaron2018}. 
However, the effect of quantum fluctuations is yet to be clarified in these works: the former considers maximally mixed states, i.e., infinite-temperature states, whereas the latter are within the mean-field approximation. In addition, it is nontrivial whether the universal FV scaling of the surface roughness occurs for far-from-equilibrium relaxation dynamics. This is important for our understanding of quantum thermalization in isolated systems, which is related to the foundation of statistical mechanics. Experiments of ultracold atoms \cite{thermalization1,thermalization2,Kinoshita2006,Hofferberth2007,Gring2012,Trotzky2012,Langen2015,Kaufman2016} have observed the thermalization processes, but such long-time universal growth of fluctuations is little known.

In this Letter, we study fluctuation growth dynamics in a 1D strongly interacting Bose-Hubbard model (BHM) from the perspective of the FV scaling of the surface roughness. We use the roughness instead of the correlation function because the roughness in the quantum system can be defined only by a local quantity and hence easy to be observed experimentally. In fact, employing analogy between fluctuating hydrodynamics and stochastic surface growth \cite{Spohn2014,Suman2014,Spohn2015,Spohn2016}, we can introduce a surface-height operator composed of local particle-number operator in the BHM. By using the surface height extended to the quantum system, we calculate the surface roughness, demonstrating the emergence of the FV scaling in the isolated quantum many-body system. All the initial states used in this work are pure states and thus our findings are obtained from the quantum dynamics triggered by purely quantum fluctuations. We have demonstrated two possibilities of the FV scaling exponents depending on the filling factor $\nu$, which are summarized in Fig.~\ref{fig1}(c). We argue that the exponents of the high-filling system follow the EW class while the low-filling (close to $1/2$) system belongs to an unconventional class. Furthermore, considering the isotropic Heisenberg spin chain as a related model, we obtain a signature of the KPZ class.

We comment on the relation between our work and previous ones on quantum transport \cite{Prosen2019,Moore2019,Ferdinand2019}. 
The previous works on transport phenomena have evaluated $z$ (which also appears in the FV scaling) using different-time correlation functions, and have discussed ballistic, diffusive, and anomalous transport. 
On the other hand, these correlation functions do not directly exhibit the FV scaling. This is in stark contrast with our roughness, which can be expressed as a sum of equal-time correlation functions as described later. To the best of our knowledge, the FV scaling shown here has, in fact, not been reported in the context of quantum transport.

{\it Theoretical model and setup.-}
We consider an $N$-boson system trapped in a 1D optical lattice, which is well described by the BHM \cite{coldgas1,stringari,pethick}. The Hamiltonian is given by 
\begin{eqnarray}
\hat{H} = -J \sum_{j=1}^{M} \Bigl(\hat{b}_{j+1}^{\dagger} \hat{b}_{j} + {\rm h.c.} \Bigl) + \frac{U}{2} \sum_{j=1}^{M}  \hat{b}_{j}^{\dagger} \hat{b}_{j}^{\dagger} \hat{b}_{j}\hat{b}_{j}, \label{BH}
\end{eqnarray}
where $\hat{b}_{j}$ and $\hat{b}_{j}^{\dagger}$ are the annihilation and creation operators at the $j$th site, respectively, $J$ is a hopping parameter, $U$ is an interaction coupling parameter, and $M$ is the number of lattice sites. We assume the periodic boundary condition.

We introduce a surface-height operator to define surface roughness in the BHM.
The key idea is the emergence of the KPZ scaling in classical fluctuating hydrodynamics, where the correlation function of the density fluctuation $\delta \rho (x,t)$ shows a similar scaling law in a correlation function of $\partial_x h(x,t)$ in the KPZ equation \cite{Spohn2014,Suman2014,Spohn2015,Spohn2016}. From this analogy, we propose the following integral quantity as a surface height in the fluctuating hydrodynamics:
\begin{eqnarray}
\int ^x \delta \rho(y,t) dy.
\label{classical_surface}
\end{eqnarray}
Extending Eq.~\eqref{classical_surface} to the 1D BHM, we introduce the following surface-height operator for the quantum discrete system: 
\begin{eqnarray}
\hat{h}_j(t)  &=& \sum_{k=1}^{j} (\hat{b}_{k}^{\dagger}(t) \hat{b}_k(t) - \nu), 
\label{height1}
\end{eqnarray}
where $\nu=N/M$ is a filling factor.
Then, the surface roughness $w_{q}(t)$ for the fluctuation of $\hat{h}_j(t)$ is defined by
\begin{eqnarray}
w_{q} (M,t) = \sqrt{ \frac{1}{M} \sum_{j=1}^{M} \biggl\langle  \Bigl(  \hat{h}_j(t)  - h_{\rm av}(t) \Bigl)^2 \biggl\rangle  }
\label{height2}
\end{eqnarray}
with the spatially averaged surface-height $h_{\rm av}(t) =  \sum_{j=1}^{M} \langle  \hat{h}_j(t) \rangle/M$.
Here, the bracket means a quantum average with an initial state. 
This roughness can be expressed as a summation of the equal-time correlation functions for the particle-fluctuation operator $\hat{b}_{k}^{\dagger}(t) \hat{b}_k(t) - \nu$.
As far as we know, none of the works have introduced the surface-height operator.

In what follows, we consider the BHM with the strong repulsive interaction ($J \ll \nu U$), which allows one to truncate the local Fock states into a few ones because the fluctuations of the particle number should be suppressed. Then, the BHM can be effectively described by spin models depending on the filling factor $\nu$. Below, we investigate the roughness dynamics for high-filling ($\nu \gg 1$) and low-filling ($\nu<1$) cases. The high-filling case can be solved by means of the SU(3) truncated Wigner approximation (TWA) method if $\nu$ satisfies $\nu J\ll U$, whereas the low-filling is exactly solvable using the Jordan-Wigner transformation.

{\it Results (high-filling case).-}
We study the surface roughness growth when the filling factor $\nu$ is much higher than unity. 
Due to the strong interaction, we can truncate the local Fock states into $|\nu+1 \rangle$, $|\nu \rangle$, $|\nu-1 \rangle$.  
Employing the truncated states, we can rewrite the original Hamiltonian \eqref{BH} as the effective spin-1 model \cite{Altman2002,Altman2007,Nagao2016,Nagao2018}:
\begin{eqnarray}
\hat{H}_{\rm S1} = - \nu J \sum_{j=1}^{M} \Big( \hat{S}_{j+1}^{x} \hat{S}_{j}^{x} + \hat{S}_{j+1}^{y} \hat{S}_{j}^{y} \Big)  +  \frac{U}{2}\sum_{j=1}^M \big( \hat{S}_{j}^{z} \big)^2
\label{EBH}
\end{eqnarray}
with the spin-1 operator $\hat{S}_{j}^{\mu}~(\mu=x,y,z)$. 
The derivation of this model is given in Supplementary material (SM) \cite{SM}.
In the spin representation, the particle-number fluctuation at the $j$th site is expressed by $\hat{S}_{j}^{z}$, and thus the surface-height operator \eqref{height1} reads $\hat{h}_j(t)  = \sum_{k=1}^{j} \hat{S}_{k}^{z}(t)$. In a restricted solid-on-solid model \cite{Nijs1989}, the same surface height is introduced for mapping from the model to a quantum spin chain in an imaginary-time formalism.

\begin{figure}[t]
\begin{center}
\includegraphics[keepaspectratio, width=8.8cm,clip]{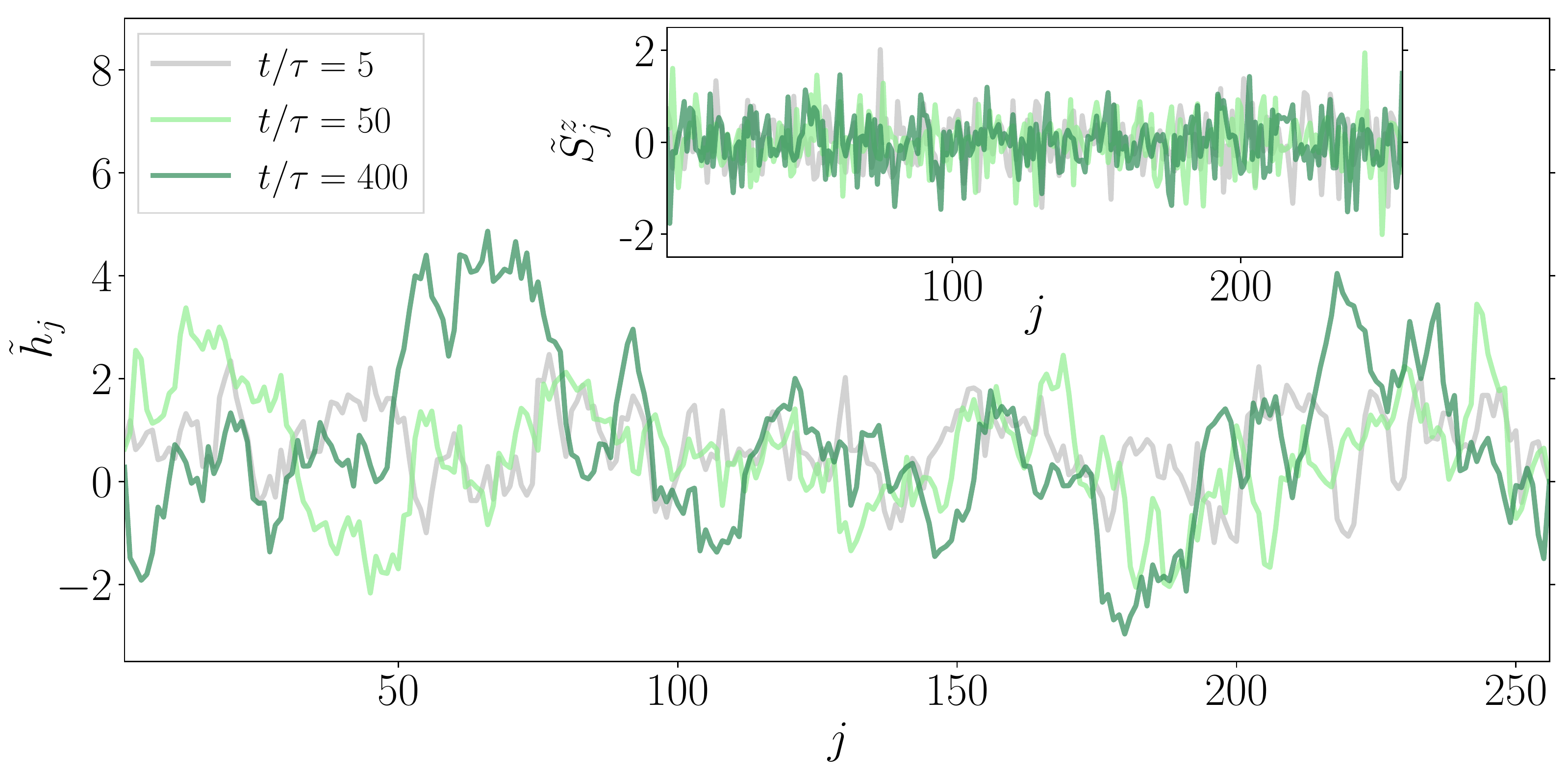}
\caption{Snapshots for the Weyl representations of (main panel) $\hat{h}_j$ and (inset) $\hat{S}_j^z$ at different time in a single trajectory of the TWA calculation with $\nu J/U=0.1$. We denote these representations by $\tilde{S}_{j}^z$ and $\tilde{h}_{j}$. The time is normalized by $\tau=\hbar/\nu J$. While the spin distributions exhibit no significant structures by eye, the distributions of the surface height clearly grow with the large-scale fluctuations.    
\label{sun_s} }
\end{center}
\end{figure}

The surface-hight distribution is constructed by the mapping rule that the eigenvalues $1,0$, and $-1$ of $\hat{S}^z_j$ are assigned to diagonally upward, horizontal, and diagonally downward lines, respectively \cite{SM}. This kind of mapping is originally developed in the simple exclusion processes \cite{Spohn1991,Sasamoto_ASEP_2005,Borodin2007,Borodin2008}, where the surface height is given by a time integral of currents. Defining the current operator $\hat{I}_j = iJ (\hat{b}_{j}^{\dagger} \hat{b}_{j-1} - \hat{b}_{j-1}^{\dagger} \hat{b}_j)/\hbar $, we can derive a similar relation for the BHM: $\hat{h}_{j}(t) =  \int_{0}^{t}  (\hat{I}_{0}(t_1) - \hat{I}_{j+1}(t_1)) dt_1$, which is obtained by integrating the Heisenberg equation for $\hat{h}_j$. This is almost same as the surface height in the simple exclusion processes except for $\hat{I}_{0}$, and suggests that there is a closer connection between the quantum roughness dynamics and the classical processes. 

We numerically solve the Schr$\rm \ddot{o}$dinger equation with \eqref{EBH} using the SU(3) TWA method \cite{Davidson2015}, and calculate the surface roughness $w_{q}(M,t)$. The numerical method works well under the condition $\nu J \ll U$, which is confirmed in SM by comparing with the exact numerical result \cite{SM}. 
The initial condition is chosen as the Mott state:
\begin{eqnarray}
| \psi_{\rm ini} \rangle = \prod_{j=1}^{M} \frac{1}{\sqrt{\nu !}} (\hat{b}_{j}^{\dagger})^{\nu}   |0 \rangle
\label{XX3}
\end{eqnarray}
with the vacuum $|0 \rangle$.

Figure~\ref{sun_s} shows the snapshots of $\hat{S}_{j}^z$ and $\hat{h}_{j}$ at different times in a single trajectory of the TWA calculation, from which we find that the surface-height distributions show clear growth of the large-scale fluctuations in time, whereas the spin distributions (particle-number fluctuations) do not grow. The surface-height dynamics looks similar to stochastic surface growth in classical models.  

\begin{figure}[t]
\begin{center}
\includegraphics[keepaspectratio, width=8.8cm,clip]{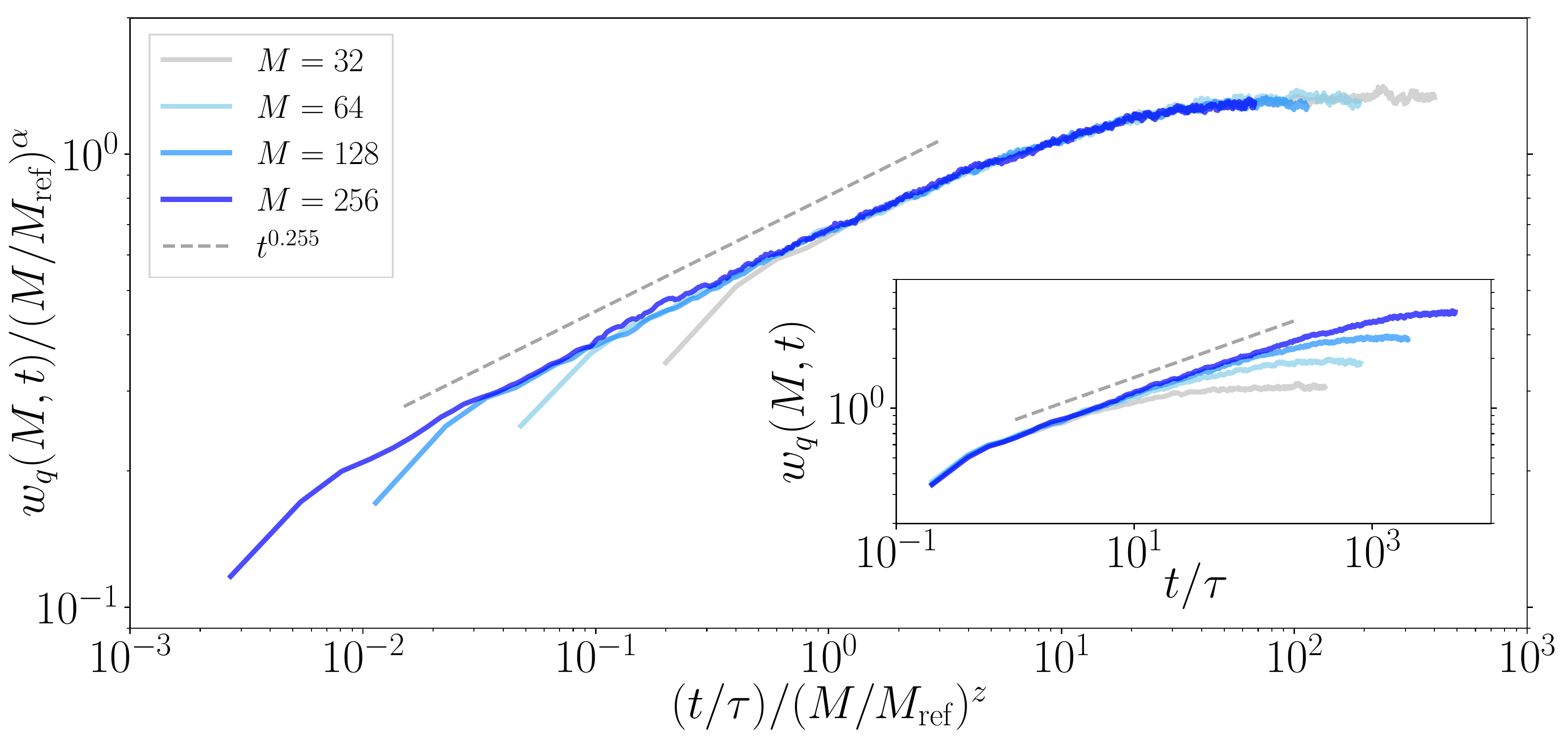}
\caption{Time evolution of $w_{q}(M,t)$ for the effective spin-1 Hamiltonian \eqref{EBH} with $\nu J/U=0.1$. The system sizes are $M=32,64,128$, and $256$, and the curves are obtained using 1000 samples. The ordinate and the abscissa are normalized by $(M/M_{\rm ref})^{\alpha}$ and $(M/M_{\rm ref})^z$. All the curves show the FV scaling with the extracted exponents $\alpha = 0.517 \pm 0.030$, $\beta = 0.255 \pm 0.012$, and $z = 2.07 \pm 0.20$. The way to extract them is described in SM \cite{SM}. The roughness $w_{q}(M,t)$ obeys the power law growth and the exponent is close to $1/4$. (Inset) Raw numerical data calculated by the SU(3) TWA method.
\label{sun_w} }
\end{center}
\end{figure}

We numerically calculate $w_{q}(M,t)$ and find that the roughness grows with increasing $t$ and $M$ as shown in Fig.~\ref{sun_w}(a). The FV scaling is expressed by $w_{q}(M,t) = s^{-\alpha} w_{q}(sM, s^{z}t)$. Substituting $s=M_{\rm ref}/M$ with the reference system size $M_{\rm ref}=32$ into this scaling relation, we obtain 
\begin{eqnarray}
w_{q}(M,t) = ( M/M_{\rm ref})^{\alpha} w_{q}(M_{\rm ref}, t (M/M_{\rm ref})^{-z}). 
\label{FV_dyna}
\end{eqnarray}
Normalizing the ordinate and the abscissa in Fig.~\ref{sun_w}(a) by $(M/M_{\rm ref})^{\alpha}$ and $(M/M_{\rm ref})^z$, respectively, we find that the curves for different system sizes collapse to a single function as in Fig.~\ref{sun_w}(b), which is the definite hallmark of the FV scaling. The extracted exponents are given by $(\alpha, \beta, z) = (0.517 \pm 0.030,  0.255 \pm 0.012, 2.07 \pm 0.20)$, which are almost identical to the exponents of the EW class (see Fig.~\ref{fig1}(c)). Here, we obtain the values of the exponents by using Eq.~\eqref{FV_dyna} and fitting the numerical data for $2<t/\tau<80$ to $ct^{\beta}$ with a constant $c$. The details of extracting the exponents are given in SM \cite{SM}. Here, we emphasize that the roughness growth in isolated quantum systems free from noises is nontrivial because, in classical systems, stochastic noises play an essential role. For example, when the noise term is absent, the classical EW model becomes a diffusive equation, which does not show the roughness growth.

We also investigate the dependence of the dynamics on the initial state, and still find the same power-law growth, details of which are given in SM \cite{SM}. 
Such a FV scaling is not found for the fluctuations of $\hat{S}^z_j$. 

Note that the initial dynamics ($t/\tau \lesssim 0.5$) shows the different type of growth. We expect that this regime strongly depends on the initial state and is non-universal because the time scale is shorter than the hopping time $\tau$. Thus, it is natural that the data in the region do not obey the FV scaling. 

Next, we consider dynamics under the condition $\nu J \simeq U$, in which the SU(3) TWA calculation is not valid. In small systems, we find a signature of the power-law growth in the roughness. However, because the time region having the power-law-like behavior is short, we cannot confirm the clear FV scaling. This result is described in SM \cite{SM}. 

\begin{figure}[t]
\begin{center}
\includegraphics[keepaspectratio, width=8.8cm,clip]{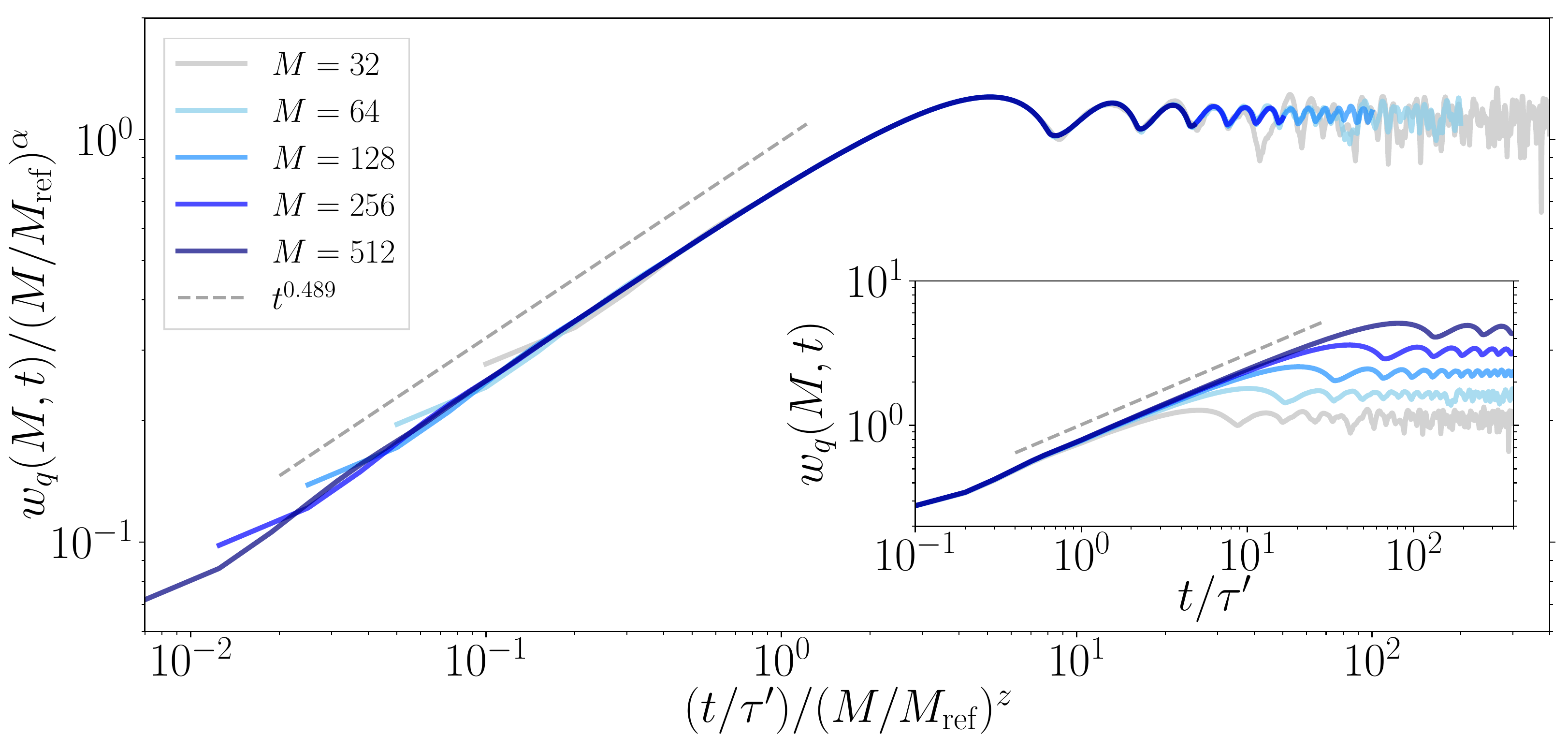}
\caption{Time evolution of $w_{q}(M,t)$ in the XX model \eqref{TXX1} starting from the staggered initial state \eqref{TXX3}. 
The ordinate and the abscissa are normalized by $(M/M_{\rm{ref}})^{\alpha}$ and $(M/M_{\rm{ref}})^z$. 
The surface roughness shows the power-law growth up to the saturation time, after which small oscillations emerge. 
Except the very early stage, the growth of the surface roughness shows the FV scaling. The extracted exponents are  $\alpha = 0.500 \pm 0.003$, $\beta = 0.489 \pm 0.004$, and $z = 1.00 \pm 0.01$. (Inset) Raw data calculated by the exact solution. The time is normalized by $\tau' = \hbar/2J$. 
\label{txx_w} }
\end{center}
\end{figure}

{\it Results (low-filling case).-}
We consider the BHM with the half filling $\nu=1/2$. 
Owing to the strong repulsive interaction, the bases of the local Fock states can be reduced to $|0\rangle$ and $|1\rangle$. As a result, the Hamiltonian~\eqref{BH} becomes the XX model \cite{sachdev2007quantum}:
\begin{eqnarray}
\hat{H}_{\rm XX} = -2J \sum_{j=1}^{M} \biggl( \hat{s}_{j+1}^x \hat{s}_{j}^x +  \hat{s}_{j+1}^y \hat{s}_{j}^y \biggl)  + ~{\rm const.} 
\label{TXX1}
\end{eqnarray}
The spin-$1/2$ operators $\hat{s}_{j}^\alpha ~(\alpha=x,y,z)$ are given by $\hat{s}_{j}^x = \bigl( \hat{b}_j^{\dagger} + \hat{b}_j \bigl)/2$, $\hat{s}_{j}^y = -i \bigl( \hat{b}_j^{\dagger} - \hat{b}_j \bigl)/2$, and $\hat{s}_{j}^z =  \hat{b}_j^{\dagger}\hat{b}_j - 1/2$, which satisfy the commutation relation $[  \hat{s}_{i}^{\alpha}, \hat{s}_{j}^{\beta} ] = i \delta_{ij} \sum_{\gamma} \epsilon_{\alpha \beta \gamma} \hat{s}_{j}^{\gamma}$. The particle-number fluctuation is given by $\hat{s}^z_j$, and thus the surface-height operator \eqref{height1} reduces to $\hat{h}_j = \sum_{k=1}^{j} \hat{s}^z_k$. Similarly to the high-filling case, we can construct the surface-height distribution by assigning an up (down) spin to a diagonally upward (downward) line \cite{SM}.
As an initial state, we use a staggered state given by
\begin{eqnarray}
| \psi_{\rm ini}\rangle = \prod_{j=1}^{M/2} \hat{b}_{2j-1}^{\dagger}   |0 \rangle, 
\label{TXX3}
\end{eqnarray}
where $M$ is assumed to be even. Under this setup, we exactly solve the Heisenberg equation with Eq.~\eqref{TXX1} by employing the Jordan-Wigner transformation \cite{sachdev2007quantum}, and calculate the exact time evolution of the surface roughness \eqref{height2}. The detail of the algebraic calculations is given in SM \cite{SM}. 

Figure~\ref{txx_w} shows the exact time evolution of $w_{q}(M,t)$ for different system sizes, which demonstrates growth of the surface roughness. Normalizing the ordinate and the abscissa in a similar manner to Eq.~\eqref{FV_dyna}, we find that all the different curves collapse to a single function except for the very early and late stages of the dynamics. The extracted power-law exponents are $(\alpha, \beta, z) = (0.500 \pm 0.003, 0.489 \pm 0.004, 1.00 \pm 0.01)$. We also investigate the dependence of the exponents on the filling factor $\nu$ and confirm that the similar exponents emerge in the late dynamics unless $\nu$ is far from $1/2$ as described in SM \cite{SM}. As far as we know, any classical models do not have these exponents, which suggests that this system belongs to an unconventional universality class. 

Here, we particularly consider the specific staggered state~\eqref{TXX3} as an initial condition because the initial roughness should be small. 
We also calculate for other initial states and still find that the FV scaling robustly appears as long as the initial roughness is small enough and the filling fraction is not too small \cite{SM}.

\begin{figure}[b]
\begin{center}
\includegraphics[keepaspectratio, width=8.8cm,clip]{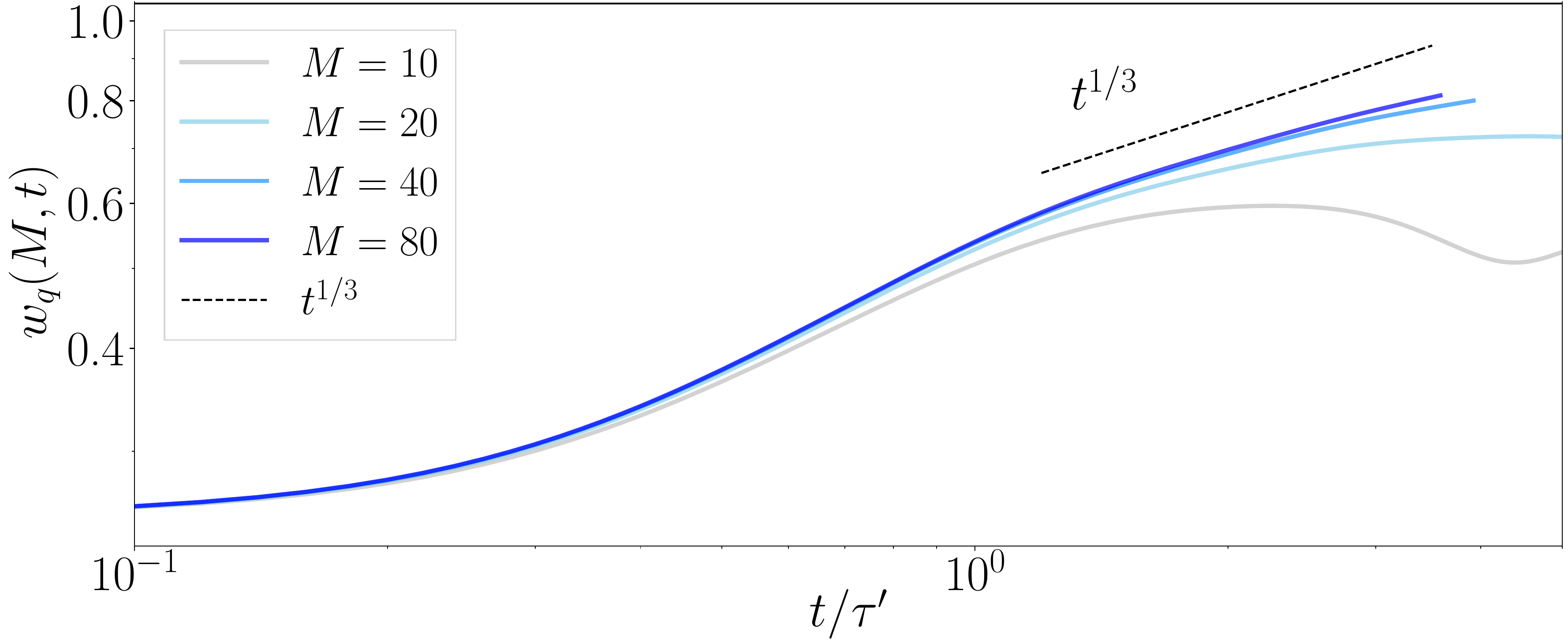}
\caption{
Time evolution of $w_{q}(M,t)$ in the isotropic Heisenberg model \eqref{IH} starting from the staggered initial state. When $M$ is larger than $40$, the roughness growth shows a signature of $1/3$ power law. We perform the calculations using the ITensor Library \cite{ITensor}.
\label{IH_w} }
\end{center}
\end{figure}

{\it Discussion.-}
As summarized in Fig.~\ref{fig1}(c), the exponent $\alpha=1/2$ seems to be model-independent. It can be analytically derived using the eigenstate thermalization hypothesis (ETH) \cite{ETH1,ETH2,Luca_2016,Mori_2018} and the cluster decomposition: $\lim_{t \rightarrow \infty} w_{q}^2(M,t) \simeq C( M +1)/2 $, where $C$ is the constant. Thus, in a large system, we obtain $\alpha=1/2$. The details of the derivation is given in SM \cite{SM}.
Note that this argument itself cannot explain $\alpha=1/2$ in the XX model since the derivation is based on ETH, which is not valid in the free-fermion model. However, the essence of the derivation is the translational invariance and no long range order, and thus we derive $\alpha=1/2$ even in the XX model if the two assumptions are valid \cite{SM}. As for the exponent $\beta$, we have not analytically obtained the value and leave it in the future work. 

Next, we discuss the KPZ class in the isotropic Heisenberg model, which is a simple extension of the XX model~\eqref{TXX1}. 
While this model is outside the framework of the BHM, it appears in many condensed-matter context and is a prototypical model for statistical mechanics. The Hamiltonian is given by 
\begin{eqnarray}
\hat{H}_{\rm IH} = -2J \sum_{j=1}^{M} \biggl( \hat{s}_{j+1}^x \hat{s}_{j}^x +  \hat{s}_{j+1}^y \hat{s}_{j}^y +  \hat{s}_{j+1}^z \hat{s}_{j}^z \biggl).
\label{IH}
\end{eqnarray}
We numerically solve the model using the matrix product state technique, and then find a signature of the KPZ class as shown in Fig.~\ref{IH_w}, which shows the time evolution of the roughness calculated by the surface-height operator $\hat{h}_j = \sum_{k=1}^{j} \hat{s}^z_k$. The roughness growth obeys a power-law-like behavior with $\beta=1/3$ in the time region $[1,3]$ where the results for $M=40$ and $80$ overlap. We leave it a future study to confirm the exponents for larger system sizes and longer timescales.

Finally, we discuss possible experiments for observing the FV scaling. The surface-height operator~\eqref{height1} is the summation of the local particle-number operator at a simultaneous time. Thus, in the low-filling case, the observation of the roughness is easier than that of spatio-temporal correlation functions by utilizing quantum gas microscopes. Experiments in ultracold atomic gases have already observed the thermalization processes in the low-filling case starting from the staggered state \cite{Trotzky2012}, and thus the FV scaling in Fig.~\ref{txx_w} can be detectable. 
Another promising testbed is a Rydberg system, in which the XX model is realizable in a highly controlled manner and the particle number fluctuations can be observed \cite{R1,R2}.
On the other hand, in the high-filling case, current experiments may not have adequate resolution for detecting one particle dynamics, and thus it may be challenging to observe the FV scaling.

{\it Conclusions and prospects.-}
We have theoretically studied the surface roughness dynamics in the strongly interacting 1D Bose gas by introducing the surface-height operator in the BHM, and then have demonstrated the emergence of the FV scaling in an isolated quantum many-body system. The extracted exponents in the high-filling case correspond to the EW class while in the low-filling case the exponents are found to be unconventional with no corresponding classical models. 

As future works, it is interesting to consider quantum thermalization in isolated systems from the viewpoint of the surface-roughness growth. The relation between simple exclusion processes and the quantum roughness growth can open an interesting avenue for connecting quantum thermalization dynamics and classical stochastic processes. It is also interesting to investigate the relation with universal dynamics for certain nonlocal quantities such as entanglement entropy and operator spreading~\cite{entanglement1,entanglement2,entanglement3,entanglement4}, which are in stark contrast to our finding because the surface-height operator is a summation of local operators. As another direction, it is important to pursuit further connections between classical and quantum roughness growth by focusing on higher-order cumulants of surface fluctuations, which may reveal the Tracy-Widom random matrix universality characteristic of the KPZ classes~\cite{Prahofer2000,Prahofer2004,sasamoto2010,Calabrese2011,Takeuchi2010,Takeuchi2011,Takeuchi2012}.

\begin{acknowledgments}
We would like to thank Ryo Hanai and Hosho Katsura for fruitful discussions.
This work was supported by JST-CREST (Grant No. JPMJCR16F2), JSPS KAKENHI (Grant Nos. JP15K17726, JP18K03538, JP19H01824, and JP19K14628), and the Program for Fostering Researchers for the Next Generation (IAR, Nagoya University) and Building of Consortia for the Development of Human Resources in Science and Technology (MEXT). R. H. was supported by the Japan Society for the Promotion of Science through Program for Leading Graduate Schools (ALPS) and JSPS fellowship (JSPS KAKENHI Grant No. JP17J03189).

\end{acknowledgments}

\bibliography{reference}

\widetext
\clearpage

\setcounter{equation}{0}
\setcounter{figure}{0}
\setcounter{section}{0}
\renewcommand{\theequation}{S-\arabic{equation}}
\renewcommand{\thefigure}{S-\arabic{figure}}

\section*{Supplemental Material for ``Family-Vicsek Scaling of Roughness Growth in a Strongly Interacting Bose Gas''}

This document describes the following topics:
\begin{itemize}
\item[  ]{ (I) derivation of the effective spin-1 model, } 
\item[  ]{ (II) SU(3) TWA method in the effective spin-1 model, }
\item[  ]{ (III) roughness dynamics in the effective spin-1 model and the dependence on initial states, }
\item[  ]{  (IV) mapping from a Fock state to a surface-height distribution, }
\item[  ]{ (V) derivation of the exponent $\alpha=1/2$ in the effective spin-1 model,}
\item[  ]{ (VI) exact solution of the surface roughness in the XX model,}
\item[  ]{ (VII) method for extracting the exponents in the surface roughness growth.}
\end{itemize}

\section{Derivation of the effective spin-1 model}
We derive the effective spin-1 model Eq.~(6) from the Bose-Hubbard model (BHM) (2) in the main text, which was originally obtained by Altman and Auerbach \cite{Altman2002}. 
The situation considered here is that the filling factor $\nu$ is much higher than unity and that the interaction energy $\nu U$ is much larger than $J$. Due to the latter assumption, we safely assume that the density fluctuations are strongly suppressed, and thus can truncate the local Fock state at a site $j$ into three states $ | \nu - 1 \rangle_j$, $| \nu  \rangle_j$, and $| \nu + 1 \rangle_j$. Employing the truncated Hilbert space, we can derive the effective spin-1 model. In the following, we follow Ref.~\cite{Nagao2016}.

First, we introduce the following bosonic operators $\hat{a}_{m,j}$ and $\hat{a}_{m,j}^{\dagger}~(m=1,0,-1)$:
\begin{eqnarray}
| \nu + m \rangle_j =  \hat{a}_{m,j}^{\dagger} | 0 \rangle_j, 
\end{eqnarray}
\begin{eqnarray}
_{j}\langle \nu + m | = ~_{j}\langle 0 | \hat{a}_{m,j},
\end{eqnarray}
\begin{eqnarray}
\sum_{m=-1}^{1} \hat{a}_{m,j}^{\dagger} \hat{a}_{m,j} = 1.
\end{eqnarray}
The last equation gives a constraint that excludes multiplied states, e.g., $\hat{a}_{m,j}^{\dagger} \hat{a}_{p,j}^{\dagger} | 0 \rangle_j$.
Then, the original creation operator $\hat{b}_{j}^{\dagger}$ of a boson can be expressed by
\begin{eqnarray}
\hat{b}_{j}^{\dagger} = \sum_{m=-1}^{1} \sqrt{\nu+m+1} \hat{a}_{m+1,j}^{\dagger} \hat{a}_{m,j}
\label{flavor_boson1}
\end{eqnarray}
with the total particle number $N_{\rm tot}$. The operator for the particle number at the $j$th site is given by
\begin{eqnarray}
\hat{\rho}_{j}= \sum_{m=-1}^{1} (\nu+m) \hat{a}_{m,j}^{\dagger} \hat{a}_{m,j}.
\end{eqnarray}

Second, we use the bosonic operators $\hat{a}_{m,j}$ and $\hat{a}_{m,j}^{\dagger}$ to transform the BHM to the effective spin-1 model.
Due to the high filling factor ($\nu \gg 1$), $\hat{b}_{j}^{\dagger}$ can be simplified to 
\begin{eqnarray}
\hat{b}_{j}^{\dagger} &\simeq& \sqrt{\nu} \bigl(  \hat{a}_{1,j}^{\dagger} \hat{a}_{0,j} + \hat{a}_{0,j}^{\dagger} \hat{a}_{-1,j} \bigl) \\
&=& \sqrt{\frac{\nu}{2}} \hat{S}_{j}^{+}, 
\label{flavor_boson2}
\end{eqnarray}
where we define the spin raising operator $\hat{S}_{j}^{+}$. The other spin operators are defined by
\begin{eqnarray}
\hat{S}_{j}^{-}  :=  \sqrt{2}(\hat{a}_{0,j}^{\dagger} \hat{a}_{1,j} + \hat{a}_{-1,j}^{\dagger} \hat{a}_{0,j}),
\end{eqnarray}
\begin{eqnarray}
\hat{S}_z := \hat{a}_{1,j}^{\dagger}  \hat{a}_{1,j} - \hat{a}_{-1,j}^{\dagger}  \hat{a}_{-1,j}.
\label{flavor_boson3}
\end{eqnarray}
The spin operators satisfy the angular-momentum commutation relation. 
Under the same approximation, the particle-number operator reads 
\begin{eqnarray}
\hat{\rho}_{j} - \nu &\simeq& \hat{a}_{1,j}^{\dagger} \hat{a}_{1,j} - \hat{a}_{-1,j}^{\dagger} \hat{a}_{-1,j} \\
&=& \hat{S}^{z}_j. 
\label{flavor_boson4}
\end{eqnarray}
Substituting all the above operators \eqref{flavor_boson2} -- \eqref{flavor_boson4} into the BHM (2) in the main text, we obtain
\begin{eqnarray}
\hat{H} &=& -J \sum_{j=1}^{M} \Bigl(\hat{b}_{j+1}^{\dagger} \hat{b}_{j} + {\rm h.c.} \Bigl) + \frac{U}{2} \sum_{j=1}^{M}  \hat{b}_{j}^{\dagger} \hat{b}_{j}^{\dagger} \hat{b}_{j}\hat{b}_{j}, \\
&=& -\frac{\nu J}{2} \sum_{j=1}^{M} \Bigl(\hat{S}_{j+1}^{+} \hat{S}_{j}^{-} + {\rm h.c.} \Bigl) + \frac{U}{2} \sum_{j=1}^{M}  \Bigl(\nu+\hat{S}^{z}_j - 1 \Bigl) \Bigl(\nu+\hat{S}^{z}_j \Bigl), \\
&=& -\frac{\nu J}{2} \sum_{j=1}^{M} \Bigl(\hat{S}_{j+1}^{+} \hat{S}_{j}^{-} + {\rm h.c.} \Bigl) + \frac{U}{2} \sum_{j=1}^{M} \Big( \hat{S}^{z}_j \Big)^2 - \mu \sum_{j=1}^{M} \hat{S}^{z}_j  + {\rm const.}
\label{effective_spin1}
\end{eqnarray}
with $\mu = - \nu U + U/2$. This Hamiltonian is invariant under global spin rotation along the $z$-axis and commutable with the spatially averaged magnetization $\sum_{j=1}^{M} \hat{S}^{z}_j/M$. 
Thus, this operator gives a trivial constant, which becomes zero for the initial Mott state considered in the main text the constant. 
Therefore, we can simplify the Hamiltonian \eqref{effective_spin1} to 
\begin{eqnarray}
\hat{H}_{\rm S1} = -\frac{\nu J}{2} \sum_{j=1}^{M} \Bigl(\hat{S}_{j+1}^{+} \hat{S}_{j}^{-} + {\rm h.c.} \Bigl) + \frac{U}{2} \sum_{j=1}^{M} \Big( \hat{S}^{z}_j \Big)^2 + {\rm const.}
\label{effective_spin2}
\end{eqnarray}
In what follows, we neglect the trivial constant.

\section{SU(3) TWA method in the effective spin-1 model}

\subsection{Formulation}
Our numerical approach for the effective spin-1 model is the SU(3) truncated Wigner approximation (TWA) \cite{Davidson2015}. 
The key ingredient of the method is to use the unitary-transformed Gell-Mann matrices $(G_{\alpha})_{\beta \gamma}$, which are one of the representations for the SU(3) Lie algebra given by
\begin{eqnarray}
&&G_1 = \frac{1}{\sqrt{2}}
\begin{pmatrix}
0 & 1 & 0 \\
1 & 0 & 1 \\
0 & 1 & 0
\end{pmatrix}
,~
G_2 = \frac{i}{\sqrt{2}}
\begin{pmatrix}
0 & -1 & 0 \\
1 & 0 & -1 \\
0 & 1 & 0
\end{pmatrix}
,~
G_3 =
\begin{pmatrix}
1 & 0 & 0 \\
0 & 0 & 0 \\
0 & 0 & -1
\end{pmatrix}
,~\\
\nonumber\\
\nonumber\\
&&G_4 = \frac{i}{\sqrt{2}}
\begin{pmatrix}
0 & 0 & 1 \\
0 & 0 & 0 \\
1 & 0 & 0
\end{pmatrix}
,~
G_5 = 
\begin{pmatrix}
0 & 0 & i \\
0 & 0 & 0 \\
i & 0 & 0
\end{pmatrix}
,~
G_6 = \frac{1}{\sqrt{2}}
\begin{pmatrix}
0 & -1 & 0 \\
-1 & 0 & 1 \\
0 & 1 & 0
\end{pmatrix}
,~\\
\nonumber\\
\nonumber\\
&&G_7 = \frac{i}{\sqrt{2}}
\begin{pmatrix}
0 & 1 & 0 \\
-1 & 0 & -1 \\
0 & 1 & 0
\end{pmatrix}
,~
G_8 = \frac{1}{\sqrt{3}}
\begin{pmatrix}
-1 & 0 & 0 \\
0 & 2 & 0 \\
0 & 0 & -1
\end{pmatrix}
.
\label{S_Gell_Mann_matrix}
\end{eqnarray}
First, we define the SU(3) operators as
\begin{eqnarray}
\hat{T}_{\alpha,j} = \sum_{\beta,\gamma=-1}^{1} \hat{a}_{\beta,j}^{\dagger} (G_{\alpha})_{\beta \gamma}  \hat{a}_{\gamma,j}, 
\label{S_Gell_Mann}
\end{eqnarray}
which obeys the SU(3) commutation relation $[\hat{T}_{\alpha,j}, \hat{T}_{\beta,k}] = \delta_{jk} \sum_{\gamma=1}^{8} f_{\alpha \beta \gamma} \hat{T}_{\gamma,j}$. 
Here, $f_{\alpha \beta \gamma}$ is the structure factor for the SU(3) group, and is given by the following an anti-symmetric tensor:
\begin{eqnarray}
&& f_{123} = f_{147} = f_{165}=f_{246}=f_{257}=f_{367}=1, \\
&& f_{178}=f_{286}=\sqrt{3}, \\
&& f_{345}=2.
\end{eqnarray}
The notation of the structure factor is same as that in Ref.~\cite{Davidson2015}.
Permuting $(\alpha, \beta, \gamma)$ of $f_{\alpha \beta \gamma}$, we can obtain the remaining non-zero components. 
Second, employing these operators, we rewrite the effective spin-1 Hamiltonian \eqref{effective_spin2} as
\begin{eqnarray}
\hat{H}_{\rm S1} =  \underbrace{ -\nu J \sum_{j=1}^{M} \Big( \hat{T}_{1,j} \hat{T}_{1,j+1} + \hat{T}_{2,j} \hat{T}_{2,j+1} \Big)  }_{ {\rm quartic~form~with~the~operators}~\hat{a}_{\alpha,j} }
+ \underbrace{ \frac{U}{2}\sum_{j=1}^{M} \biggl( \frac{2}{3} - \frac{1}{\sqrt{3}}\hat{T}_{8,j} \biggl) }_{ {\rm quadratic~form~with~the~operators}~\hat{a}_{\alpha,j}}, 
 \label{S_EBH2}
\end{eqnarray}
where we have used the following relations:
\begin{eqnarray}
&&\hat{T}_{1,j} = \hat{S}_{j}^x \label{S_T1}\\
&&\hat{T}_{2,j} = \hat{S}_{j}^y \label{S_T2}\\
&&\hat{T}_{3,j} = \hat{S}_{j}^z \label{S_T3}\\
&&\hat{T}_{8,j} = \frac{2}{\sqrt{3}} - \sqrt{3} \Big( \hat{S}^z_j \Big)^2. \label{S_T4}
\label{S_relation}
\end{eqnarray}
Here, Eqs.~\eqref{S_T1}-\eqref{S_T3} are obtained by the fact that the matrices $G_1$, $G_2$, and $G_3$ are identical to the spin-1 matrices.
In addition, Eq.~\eqref{S_T4} comes from $\sqrt{3} G_8= 2 E - 3 G_3^2 $ with a $3 \times 3$ identity matrix $E$.  

When the hopping parameter $J$ is zero, we can exactly solve the Shr$\rm \ddot{o}$dinger equation of Eq.~\eqref{S_EBH2} because the Hamiltonian is quadratic in terms of the bosonic operators $\hat{a}_{\alpha,j}$ and $\hat{a}_{\alpha,j}^{\dagger}$. In this case, the TWA becomes exact. In the regime $\nu J \ll U$, the interaction term is dominant in comparison with the hopping one, and thus the quantum many-body dynamics can be well approximated by the SU(3) TWA. In the next subsection, we discuss the validity of the TWA using the comparison with exact numerical simulations. 

We explain how to apply the TWA to Eq.~\eqref{S_EBH2} \cite{Davidson2015}.
A quantum average for a physical quantity $\hat{A}$ is given by
\begin{eqnarray}
\langle \hat{A} \rangle &=& \langle \psi(t) | \hat{A} | \psi(t) \rangle \\
&=& \int \Big\{ \prod_{\alpha,i} dX_{\alpha,i}(0)  \Big\} P_{\rm W}( \{ X_{\alpha,i}(0) \})  \big[ \hat{A}(t) \big]_{\rm W}( \{ X_{\alpha,i}(0) \} ) \\
&\simeq& \int \Big\{ \prod_{\alpha,i} dX_{\alpha,i}(0)  \Big\} P_{\rm W}( \{ X_{\alpha,i}(0) \})  A_{\rm W}( \{ X_{\alpha,i}(t) \} ) \label{S_TWA1}.
\end{eqnarray}
Here, $X_{\alpha,i}(0)$, $P_{\rm W}( \{ X_{\alpha,i}(0) \})$ and $\big[ \hat{A}(t) \big]_{\rm W}( \{ X_{\alpha,i}(0) \} )$ are the Weyl representations for $\hat{T}_{\alpha,i}(0)$, $ | \psi(0) \rangle  \langle \psi(0) | $, and $\hat{A}(t)$, respectively. 
In the third line, we use the TWA, in which $A_{\rm W}( \{ X_{\alpha,i}(t) \} )$ is calculated from the Weyl representation of $\hat{A}(0)$ (denoted by $A_{\rm W}(\cdots)$) and the classical variables $X_{\alpha,i}(t)$ obeying
\begin{eqnarray}
\hbar \frac{\partial}{\partial t} X_{\alpha,j} = f_{\alpha \beta \gamma} \frac{\partial H_{\rm w}}{\partial X_{\beta,j}} X_{\gamma,j}, 
\label{EOM}
\end{eqnarray}
\begin{eqnarray}
H_{\rm w} =  - \nu  J \sum_{j=1}^{M} \Big( X_{1,j} X_{1,j+1} + X_{2,j} X_{2,j+1} \Big) + \frac{U}{2}\sum_{j=1}^{M} \biggl( \frac{2}{3} - \frac{1}{\sqrt{3}}X_{8,j} \biggl). 
\label{EBH3}
\end{eqnarray}
The concrete expressions are given by
\begin{eqnarray}
\hbar \frac{\partial}{\partial t} X_{1,j} &=& - \nu J  X_{3,j} \Big( X_{2,j+1} + X_{2,j-1} \Big) + \frac{U}{2} X_{7,j}, \label{EOM1} \\
\hbar \frac{\partial}{\partial t} X_{2,j} &=&  \nu J  X_{3,j} \Big( X_{1,j+1} + X_{1,j-1} \Big)  - \frac{U}{2} X_{6,j}, \\
\hbar \frac{\partial}{\partial t} X_{3,j} &=&  \nu J  \biggl\{ X_{1,j} \Big( X_{2,j+1} + X_{2,j-1} \Big) -  X_{2,j} \Big( X_{1,j+1} + X_{1,j-1} \Big)  \biggl\}, \\
\hbar \frac{\partial}{\partial t} X_{4,j} &=&  \nu J  \biggl\{ X_{7,j} \Big( X_{1,j+1} + X_{1,j-1} \Big) +  X_{6,j} \Big( X_{2,j+1} + X_{2,j-1} \Big)  \biggl\}, \\
\hbar \frac{\partial}{\partial t} X_{5,j} &=&  \nu J  \biggl\{ X_{7,j} \Big( X_{2,j+1} + X_{2,j-1} \Big) -  X_{6,j} \Big( X_{1,j+1} + X_{1,j-1} \Big)  \biggl\}, \\
\hbar \frac{\partial}{\partial t} X_{6,j} &=&  \nu J  \biggl\{ X_{5,j} \Big( X_{1,j+1} + X_{1,j-1} \Big) -  X_{4,j} \Big( X_{2,j+1} + X_{2,j-1} \Big) - \sqrt{3} X_{8,j} \Big( X_{2,j+1} + X_{2,j-1} \Big)   \biggl\} + \frac{U}{2} X_{2,j}, \\
\hbar \frac{\partial}{\partial t} X_{7,j} &=&  \nu J  \biggl\{   \Big( \sqrt{3} X_{8,j} - X_{4,j} \Big) \Big( X_{1,j+1} + X_{1,j-1} \Big) - X_{5,j} \Big( X_{2,j+1} + X_{2,j-1} \Big)   \biggl\} - \frac{U}{2} X_{1,j},  \\
\hbar \frac{\partial}{\partial t} X_{8,j} &=& \sqrt{3} \nu  J  \biggl\{   X_{6,j} \Big( X_{2,j+1} + X_{2,j-1} \Big) - X_{7,j} \Big( X_{1,j+1} + X_{1,j-1} \Big)   \biggl\} \label{EOM2} .
\end{eqnarray}
According to Eq.~\eqref{S_TWA1}, we sample initial classical variables $X_{j,\alpha}(0)$ from the Wigner function $P_{\rm W}( \{ X_{j,\alpha}(0) \})$, and subsequently we solve Eqs.~\eqref{EOM1}--\eqref{EOM2} and then calculate $A_{\rm W}( \{ X_{j,\alpha}(t) \} )$. Repeating this procedure many times, we take the ensemble average for $A_{\rm W}( \{ X_{j,\alpha}(t) \} )$ and then obtain $\langle \hat{A} \rangle$ in Eq.~\eqref{S_TWA1}.

Finally, we comment on the Wigner function $P_{\rm W}( \{ X_{j,\alpha}(0) \})$ that generates the initial condition.
The initial quantum state considered in the main text is the Mott state:
\begin{eqnarray}
|\psi (0) \rangle &=&  \prod_{j=1}^{M}  \frac{1}{\sqrt{\nu !}} (  \hat{b}_j^{\dagger}  )^{\nu} |0 \rangle \\
&=&  \prod_{j=1}^{M} | S^z_j=0 \rangle
\label{EBH4}
\end{eqnarray}
with the vaccum $|0 \rangle$. In general, it is difficult to take the ensemble average for the Mott state exactly and the previous literatures use the Gaussian approximation \cite{Davidson2015,Nagao2018}.
Employing the method developed in Ref.~\cite{Davidson2015}, we obtain the following Wigner function:
\begin{eqnarray}
P_{\rm W}( \{ X_{i,\alpha} \}) = \mathcal{A} \prod_{j=1}^{M}  {\rm exp}\biggl[    - \frac{1}{2} \Big( X_{1,j}^2 + X_{2,j}^2 + X_{6,j}^2 + X_{7,j}^2 \Big) \biggl] \delta(X_{3,j})\delta(X_{4,j})\delta(X_{5,j})\delta(X_{8,j}-\frac{2}{\sqrt{3}}), \label{S_EBH4}
\end{eqnarray}
where we use the notation $X_{i,\alpha}$ for $X_{i,\alpha}(0)$ for brevity and $\mathcal{A}$ is the normalization constant. 
Therefore, the initial condition for Eqs.~\eqref{EOM1}--\eqref{EOM2} is given by
\begin{eqnarray}
&& X_{1,j} = R_{1,j}, \\
&& X_{2,j} = R_{2,j}, \\
&& X_{3,j} = 0, \\
&& X_{4,j} = 0, \\
&& X_{5,j} = 0, \\
&& X_{6,j} = R_{3,j}, \\
&& X_{7,j} = R_{4,j}, \\
&& X_{8,j} = \frac{2}{\sqrt{3}}, 
\label{S_EBH4_1}
\end{eqnarray}
where $R_{\alpha,j}$ is independently sampled by the standard normal distribution. The detailed formulation of the SU(3) TWA is described in Ref.~\cite{Davidson2015}.

\begin{figure}[t]
\begin{center}
\includegraphics[keepaspectratio, width=17.9cm,clip]{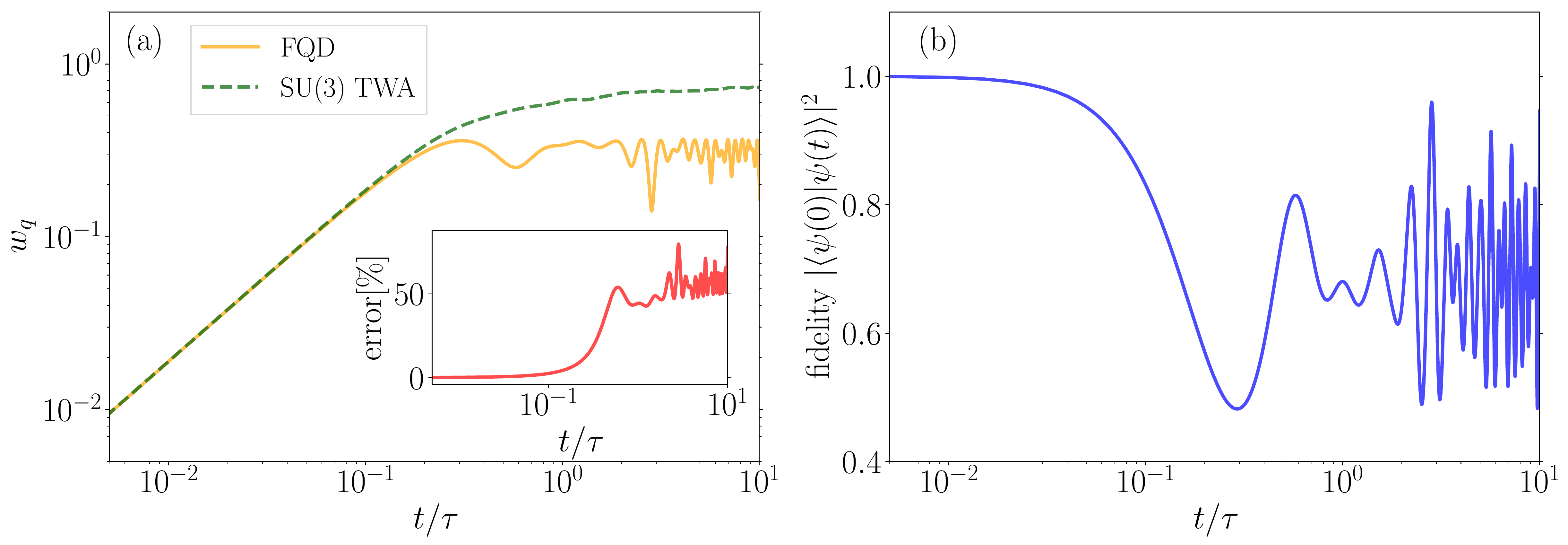}
\caption{ (a) Numerical results of the surface roughness for the SU(3) TWA method (green dashed line) and FQD (yellow solid line). 
The lattice number is $M=10$ and the other parameters satisfies $\nu J/U = 0.1$, which is same as those of Fig.~3 in the main text. The inset shows the relative difference  between the two methods. The time is normalized by $\tau = \hbar/\nu J$. (b) Time evolution of the fidelity in the FQD. In the early stage of the dynamics, the fidelity decreases in time. However, around $t/\tau \sim 0.3$, we find that the fidelity increases and then the system exhibits quantum revival over time. The TWA result begins to deviate the FQD one after the emergence of the revival. This behavior is due to small system sizes. 
\label{fqd_spin1} }
\end{center}
\end{figure}

\begin{figure}[t]
\begin{center}
\includegraphics[keepaspectratio, width=17.9cm,clip]{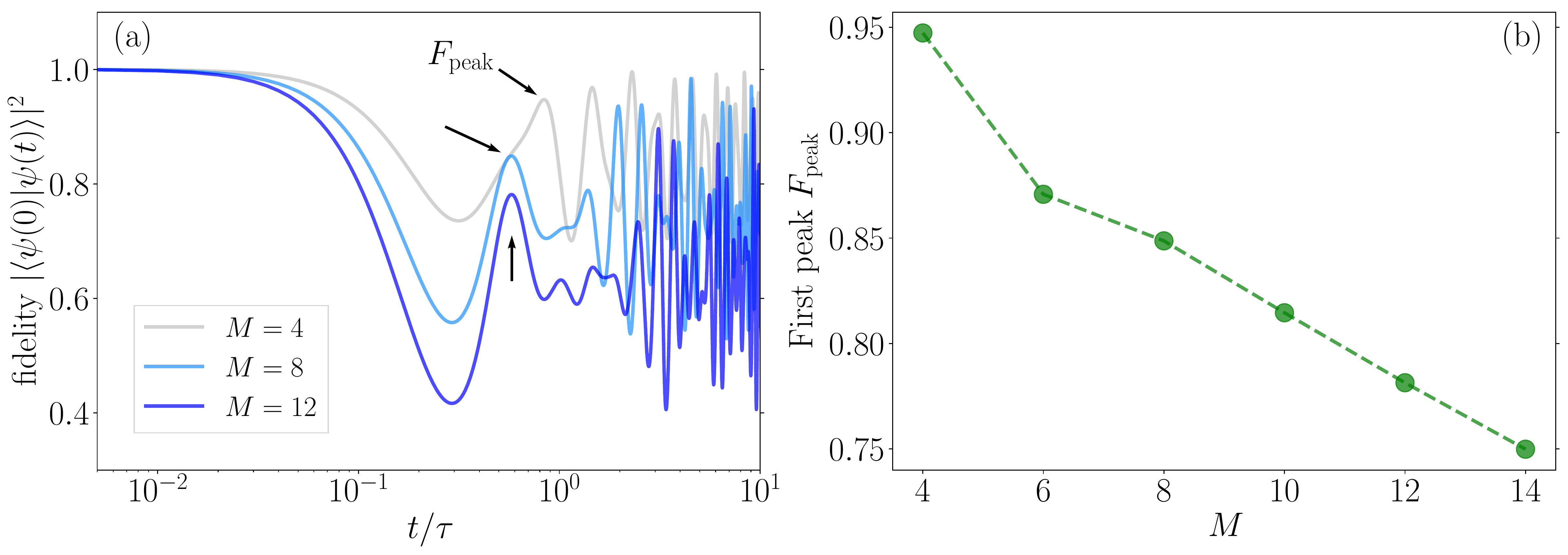}
\caption{(a) Time evolution of the fidelity for $M=4,8$, and $12$ for $\nu J/U=0.1$. As the system size becomes larger, the fidelity decreases. The arrows indicate the positions of the first peaks. 
(b) Dependence of the first-peak height $F_{\rm peak}$ on the system size $M$. As the system size increases, the peak becomes small.  
\label{fqd_spin2} }
\end{center}
\end{figure}

\subsection{Comparison with the SU(3) TWA and full quantum dynamics}
We directly solve the Schr$\rm \ddot{o}$dinger equation of Eq.~\eqref{effective_spin2}, calculating the surface roughness in full quantum dynamics (FQD), and compare it with the numerical results obtained by the SU(3) TWA. In the FQD, we numerically obtain the state vector using the Crank-Nicolson method. 

Figure~\ref{fqd_spin1}(a) shows time evolution of the surface roughness. We find that the surface roughness growths calculated by both methods show excellent agreement in the early stage of the dynamics. In the late stage where the surface roughness is saturated, however, the deviation becomes large as shown in the inset. We expect that this deviation is caused by quantum revival. In fact, as shown in Fig.~\ref{fqd_spin1}(b), fidelity $| \langle \psi(0) | \psi(t) \rangle |^2$ in the FQD shows the large oscillation without decaying, and the onset time of the revival is close to the time at which the error in the inset of Fig.~\ref{fqd_spin1} begins to increase. It is known that such revivals cannot be described by the TWA \cite{TWA1} and the validity of TWA gets better as quantum revivals are weakened.

We systematically investigate the fidelity by changing the system size $M$ as shown in Fig.~\eqref{fqd_spin2}(a), from which we find that the fidelity decreases as the system size increases. Figure~\ref{fqd_spin2}(b) plots the first-peak height $F_{\rm peak}$ of the fidelity as a function of the system size. When we extrapolate the data using a linear function, the fidelity becomes quite small in $M>60$. Thus, we expect that the quantum revival does not occur in such a large system, and that the SU(3) TWA method works well.

\begin{figure}[t]
\begin{center}
\includegraphics[keepaspectratio, width=17.9cm,clip]{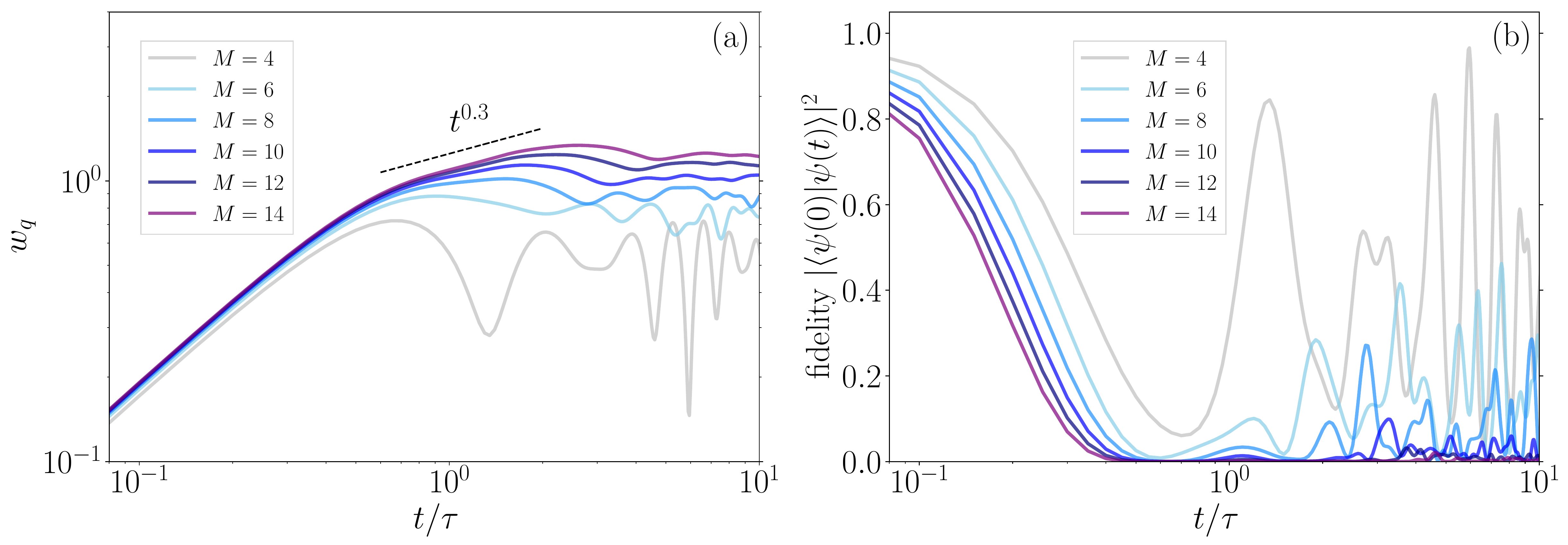}
\caption{ (a) Numerical results of the surface roughness in the FQD with $\nu J / U = 1$. The system sizes are $M=4,6,8,10,12$ and $14$. 
As the system size increases, signatures of a power-law-like growth become clearer in $t/\tau>0.6$ although the time regime is narrow. 
The dashed guide line is proportional to $t^{0.3}$. (b) Time evolution of the fidelity. When the system size is small, the fidelity shows the quantum revival. 
However, the behavior is suppressed as the system size becomes larger. 
\label{beyond_twa} }
\end{center}
\end{figure}

\subsection{Time evolution of the surface roughness beyond the SU(3) TWA}
We consider FQD for $\nu J /U \simeq 1$, in which the SU(3) TWA method does not work well. 
Figure~\ref{beyond_twa} shows the numerical results for $\nu J/U=1$ and $M=4,6,8,10,12$, and $14$. 
In Fig.~\ref{beyond_twa}(a), we find a signature of power-law-like growth in $t/\tau > 0.6$. 
However, we cannot determine the power exponents because the time regime of the power-law growth is narrow. Here, the growth at the early stage of the dynamics ($t/\tau < 0.5$) is quite similar to Fig.~3 in the main text, and we expect that this is non-universal because the time region is much smaller than the hopping time scale $\tau=\hbar/\nu J$. 

We also calculate the fidelity of the FQD in Fig.~\ref{beyond_twa}(b), and find that it rapidly decays to zero without quantum revival in the larger systems, for which the power-law-like growth of the roughness is obtained. On the other hand, the small systems show the quantum revival and do not have any signatures of the power-law-like growth. The oscillation of fidelity is correlated to that of the roughness.

This result suggests that the quantum revival breaks the power-low growth for the Family-Vicsek(FV) scaling. It also supports our argument for $\nu J \ll U$ in the previous subsection that  the SU(3) TWA becomes valid when the system size is large enough so that the quantum revival does not occur.

\section{roughness dynamics in the effective spin-1 model and the dependence on initial states}
In the main text, we consider the roughness growth dynamics starting from the specific initial state~\eqref{EBH4}. Here, we numerically investigate the dependence of the dynamics on initial states by using the following state:
\begin{eqnarray}
|\psi (0) \rangle =  \prod_{j \in \mathcal{S}_0} | S^z_j=0 \rangle \prod_{j \in \mathcal{S}_1} | S^z_j=1 \rangle, 
\label{new_initial_state}
\end{eqnarray}
where $\mathcal{S}_0$ and $\mathcal{S}_1$ are sets of the lattice points for $| S^z_j=0 \rangle$ and $| S^z_j=1 \rangle$, respectively.
We denote the number of the lattice points in $\mathcal{S}_1$ by $\triangle M$, and put the sites at equal intervals. By tuning $\triangle M$, we systematically investigate how the roughness growth changes. 
 
\begin{figure}[t]
\begin{center}
\includegraphics[keepaspectratio, width=12cm,clip]{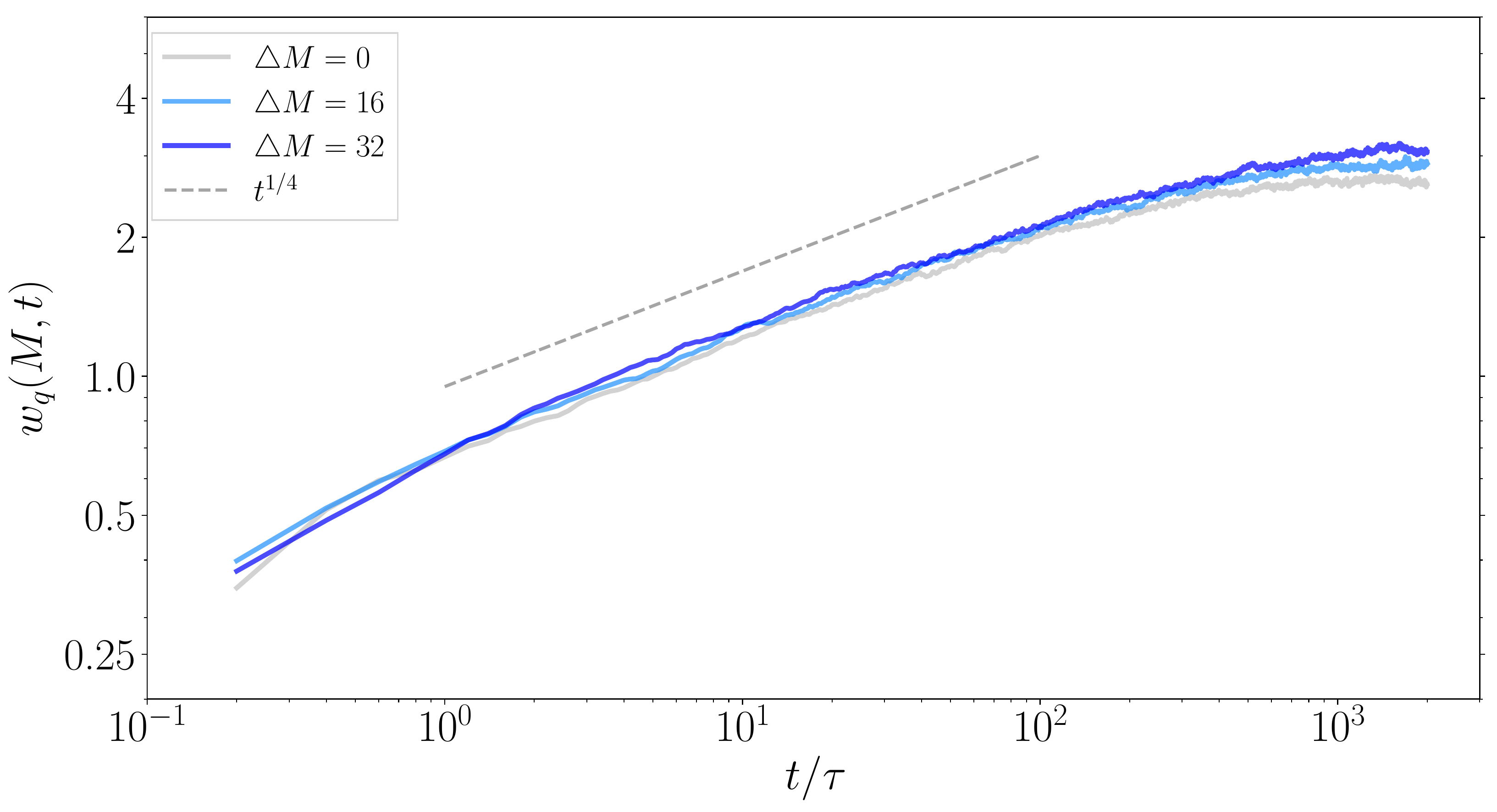}
\caption{
Time evolution of the roughness for $M=128$ and $\triangle M = 0, 16, 32$ using the initial state~\eqref{new_initial_state}. 
\label{spin1_initial_W} }
\end{center}
\end{figure}
 
To implement the SU(3) TWA calculation, we need the initial condition for $X_{\alpha,j}$ of Eqs.~\eqref{EOM1}--\eqref{EOM2}. The initial values at $\mathcal{S}_0$ are sampled by Eq.~\eqref{S_EBH4}--\eqref{S_EBH4_1}, and those at $\mathcal{S}_1$ are done by 
\begin{eqnarray}
&& X_{1,j} = \frac{R_{1,j}}{\sqrt{2}}, \\
&& X_{2,j} = \frac{R_{2,j}}{\sqrt{2}}, \\
&& X_{3,j} = 1, \\
&& X_{4,j} = R_{3,j}, \\
&& X_{5,j} = R_{4,j}, \\
&& X_{6,j} = -\frac{R_{1,j}}{\sqrt{2}}, \\
&& X_{7,j} = -\frac{R_{2,j}}{\sqrt{2}},, \\
&& X_{8,j} = -\frac{1}{\sqrt{3}}.
\label{S_EBH4_3}
\end{eqnarray}
This initial condition for the sites $\mathcal{S}_1$ can be derived using the Gaussian approximation mentioned above \cite{Davidson2015}.

Figure~\ref{spin1_initial_W} shows the roughness growth for $M=128$ and $\triangle M = 0, 16, 32$, from which we confirm that the time evolution of the roughness is not greatly affected by the initial condition.

\section{mapping from a Fock state to a surface-height distribution}
\begin{figure}[t]
\begin{center}
\includegraphics[keepaspectratio, width=14cm,clip]{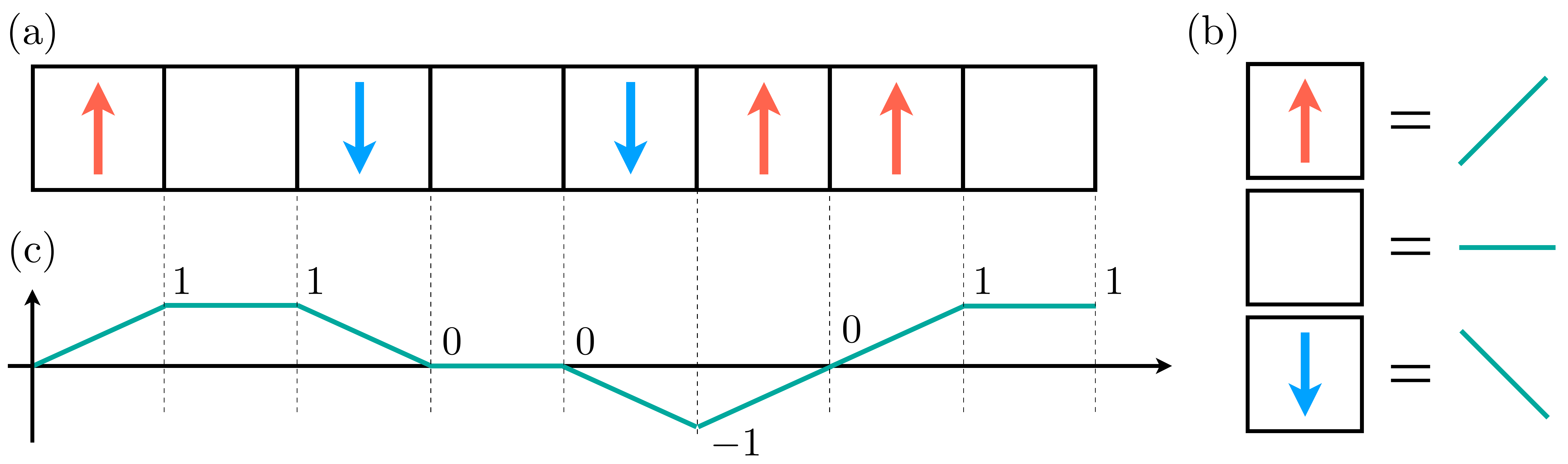}
\caption{Construction of the surface height in the spin-1 model. (a) Spin state $ |1,0,-1,0,-1,1,1,0 \rangle$. The up-arrow, vacant, and down-arrow boxes represent the eigenvalues $1$, $0$, and $-1$ of $\hat{S}_{j}^{z}$, respectively. (b) Mapping rule for the local spins. The eigenvalues $1,0$, and $-1$ are assigned to the diagonally upward, horizontal, and diagonally downward lines, respectively. (c) Surface-height distribution for (a). The numbers are the eigenvalues of the surface-height operator $\hat{h}_j$.
\label{surface_spin1} }
\end{center}
\end{figure}

We can construct the surface-height distribution using the following mapping rule when a Fock state is given. 
In the effective spin-1 model (high-filling case), we replace the eigenvalues $1$, $0$, and $-1$ of the operator $\hat{S}^z_j$ by a up-arrow, a vacant, and a down-arrow, respectively. Then, the surface-height distribution corresponding to the Fock state can be readily obtained as shown in Fig.~\ref{surface_spin1}. In the spin-1/2 XX model (low-filling case), the eigenvalues $1/2$ and $-1/2$ of the operator $\hat{s}^z_j$ are assigned to a up-arrow and a down-arrow for the construction of the surface-height distribution. 

\section{Derivation of the exponent $\alpha=1/2$ in the effective spin-1 model}
In the main text, numerically calculating the roughness in the effective spin-1 model with the initial state~\eqref{EBH4}, we find the exponent $\alpha \simeq 0.517$ in the FV scaling as shown in Fig.~3 of the main text. This section gives an analytical derivation of the exponent by employing the eigenstate thermalization hypothesis (ETH) \cite{Luca_2016,Mori_2018}. 
The essence of this calculation is translational invariance and cluster decomposition of the stationary state.

By definition, we readily obtain
\begin{eqnarray}
w_{q}^2(t,M) = \frac{1}{M} \sum_{j=1}^{M} \langle \hat{h}^2_j(t) \rangle -  h_{\rm av}^2(t).
\label{S_roughness}
\end{eqnarray}
We estimate each term in Eq.~\eqref{S_roughness}. 
Let us start with the second term
\begin{eqnarray}
h_{\rm av}(t) &=& \frac{1}{M} \sum_{j=1}^{M} \sum_{k=1}^{j} \langle \hat{S}^z_k(t) \rangle.
\label{hav1}
\end{eqnarray}
The effective spin-1 Hamiltonian~\eqref{effective_spin2} is invariant under global spin-rotation along the $z$-axis, which leads to the total spin-conservation given by
\begin{eqnarray}
\sum_{k=1}^{M} \langle \hat{S}^z_k(t) \rangle = 0, 
\label{hav2}
\end{eqnarray}
where we use the fact that the initial Mott state (7) in the main text has zero magnetization. 
Since the system respects spatial translation symmetry, we obtain $\langle \hat{S}^z_i (t) \rangle = \langle \hat{S}^z_j (t) \rangle$ for $i \neq j$. Combining this relation with Eq.~\eqref{hav2}, we find 
\begin{eqnarray}
\langle \hat{S}^z_k(t) \rangle = 0.
\label{hav3}
\end{eqnarray}
Thus, the spatial averaged surface-height becomes
\begin{eqnarray}
h_{\rm av}(t) = 0.
\label{hav4}
\end{eqnarray}
This calcualtion clearly shows that the translational invariance of the density matrix plays a crucial role in deriving this result.

As for the first term, we can approximately estimate the dependence on the system size $M$, 
\begin{eqnarray}
\lim_{t \to \infty} \frac{1}{M} \sum_{j=1}^{M} \langle \hat{h}^2_j(t) \rangle  
&=& \lim_{t \to \infty} \frac{1}{M} \sum_{j=1}^{M} \sum_{k=1}^{j} \sum_{l=1}^{j} \langle \hat{S}^z_k(t) \hat{S}^z_l(t) \rangle, \\
&\simeq&  \frac{1}{M} \sum_{j=1}^{M} \sum_{k=1}^{j} \sum_{l=1}^{j} \langle \hat{S}^z_k \hat{S}^z_l \rangle_{T} 
\label{a1}
\end{eqnarray}
where we have assumed the ETH and replace the quantum average by the thermal canonical one with the effective temperature $T$.
To evaluate the two-point spin correlation function in Eq.~\eqref{a1}, we note that a one-dimensional system with local interactions does not have any long-range orders at finite temperature, and thus safely assume cluster decomposition given by  
\begin{eqnarray}
 \langle \hat{S}^z_k \hat{S}^z_l \rangle_{T}  &\simeq&   \langle \hat{S}^z_k \rangle_{T} \langle  \hat{S}^z_l \rangle_{T} + B e^{-|k-l|/\xi}~~~(|k-l| \gg \xi) \\
 &=& B e^{-|k-l|/\xi}~~~(|k-l| \gg \xi), 
\label{a2}
\end{eqnarray}
where $B$ is a constant and $\xi$ is a coherence length which is the order of the lattice constant. Here, we have used Eq.~\eqref{hav3} to obtain the second line. This allows one to approximate the summation of the spin correlation function as follows:
\begin{eqnarray}
\sum_{l=1}^{j} \langle \hat{S}^z_k \hat{S}^z_l \rangle_{T}  &\simeq& \sum_{\substack{ l=1 \\ |k-l| \lesssim	 \xi }}^{j} \langle \hat{S}^z_k \hat{S}^z_l \rangle_{T}   +  \sum_{\substack{ l=1 \\ |k-l| \gg \xi }}^{j} \langle \hat{S}^z_k \hat{S}^z_l \rangle_{T} \\
&\simeq& \sum_{\substack{ l=1 \\ |k-l| \lesssim	 \xi }}^{j} \langle \hat{S}^z_k \hat{S}^z_l \rangle_{T} \\
&\simeq& C, 
\label{a3}
\end{eqnarray}
where $C$ is the constant whose order is $\xi$. In the second line, we neglect the exponentially small term. 
Substituting Eq.~\eqref{a3} into Eq.~\eqref{a1}, we obtain
\begin{eqnarray}
\lim_{t \to \infty} \frac{1}{M} \sum_{j=1}^{M} \langle \hat{h}^2_j(t) \rangle  
&\simeq& \frac{1}{M}\sum_{j=1}^{M} \sum_{k=1}^{j} C, \\
= \frac{C}{2} (M+1)
\label{a4}
\end{eqnarray}
From this calculation, one finds that Eq.~\eqref{a4} can be derived whenever the cluster decomposition is valid in a given density matrix. 

We substitute Eqs.~\eqref{hav4} and \eqref{a4} into Eq.~\eqref{S_roughness}, and derive
\begin{eqnarray}
\lim_{t \to \infty} w_{q}^2(t,M)  
\simeq\frac{C}{2} (M+1).
\label{a5}
\end{eqnarray}
Thus, in a large system, we obtain $\alpha=1/2$, which is close to $\alpha \simeq 0.517$ in the effective spin-1 model with the initial state~\eqref{EBH4},

Finally, we comment on the exponent of the integrable XX model. Since this model breaks the ETH, the above discussion cannot be directly applied. On the other hand, if we assume that the stationary state, which is believed to be described by the generalized Gibbs ensemble, satisfies the cluster decomposition principle and the translation invariance, $\alpha=1/2$ is derived similarly.

\section{Exact solution of the surface roughness in the XX model}
Using the XX model (9) in the main text, we give an analytical expression of the surface roughness $w_{q}(M,t)$ with a filling factor $\nu$ ($0<\nu<1$). 

\subsection{Jordan-Wigner transformation}
The XX model can be exactly solved using the Jordan-Wigner transformation defined by
\begin{eqnarray}
&&\hat{s}_{j}^{-}   =   {\rm exp}\biggl(  -i \pi \sum_{k=1}^{j-1} \hat{f}_{k}^{\dagger} \hat{f}_{k} \biggl) \hat{f}_{j}, \\
&&\hat{s}_{j}^{+}  =   {\rm exp}\biggl(  i \pi \sum_{k=1}^{j-1} \hat{f}_{k}^{\dagger} \hat{f}_{k} \biggl) \hat{f}_{j}^{\dagger} \\
&&\hat{s}_{j}^z    =   \hat{f}_{j}^{\dagger}  \hat{f}_{j} - \frac{1}{2}, 
\label{JW1}
\end{eqnarray}
where $\hat{f}_{j}$ and $\hat{f}_{j}^{\dagger}$ are annihilation and creation fermionic operators.
The Jordan-Wigner transformation maps the XX Hamiltonian (9) of the main text into 
\begin{eqnarray}
\hat{H}_{\rm XX} = -A \sum_{j=1}^{M} \biggl( \hat{f}_{j+1}^{\dagger} \hat{f}_j + \hat{f}_{j}^{\dagger} \hat{f}_{j+1} \biggl) + {\rm const.}
\label{JW2}
\end{eqnarray}
with a constant $A = 2J$. 
Thus, the original Hamiltonian becomes the free fermion model, for which we can exactly obtain the eigenvalues and eigenstates.

\subsection{Exact diagonalization}
We diagonalize Eq.~\eqref{JW2} with the periodic boundary condition and assume that the lattice number $M$ is even.
The annihilation operator can be expanded as
\begin{eqnarray}
&& \hat{f}_j = \frac{1}{\sqrt{M}} \sum_{\alpha \in \Omega} \hat{d}_{\alpha} {\rm e}^{ i \alpha j }, \label{ES1} \\
&& \Omega = \biggl\{ \pm \frac{\pi}{M}, \pm \frac{3\pi}{M}, \cdots, \pm \frac{\pi(M-1)}{M}  \biggl\}
\end{eqnarray} 
with a fermionic operator $\hat{d}_{\alpha}$ in the wavenumber space. Substituting Eq.~\eqref{ES1} into Eq.~\eqref{JW2}, we obtain
\begin{eqnarray}
&& \hat{H}_{\rm XX} = \sum_{\alpha \in \Omega} E_{\alpha}  \hat{d}_{\alpha}^{\dagger} \hat{d}_{\alpha} + {\rm const.}, \\
&& E_{\alpha} = - 2A {\rm cos}( \alpha ).
\label{ES2}
\end{eqnarray}

\subsection{Expression of the surface-height operator in terms of the fermions}
We express the density fluctuation $ \delta \hat{\rho}_j$ in terms of the fermionic operators:
\begin{eqnarray}
\delta \hat{\rho}_j &=&  \hat{b}_{j}^{\dagger} \hat{b}_{j} - \nu ~~~~~~~~~({\rm boson~representation}) \\
&=&  \hat{s}_{j}^z + \frac{1}{2} - \nu ~~~~~({\rm spin~representation}) \\
&=&  \hat{f}_{j}^{\dagger} \hat{f}_{j}  - \nu ~~~~~~~~({\rm fermion~representation}).
\label{R1}
\end{eqnarray}
Then, the surface-height operator is given by
\begin{eqnarray}
\hat{h}_j(t) &=& \sum_{k=1}^{j} \delta\hat{\rho}_k(t),  \\
&=& \sum_{k=1}^{j} \hat{f}_{k}^{\dagger} \hat{f}_{k}  - \nu j. 
\label{R2}
\end{eqnarray}

\subsection{Exact time evolution of the surface roughness}
Employing the exact eigenenergy \eqref{ES2} and the transformation \eqref{ES1}, we calculate time evolution of the roughness (5) using Eq.~\eqref{S_roughness}.
In what follows, we use the Heisenberg representation and obtain
\begin{eqnarray}
\hat{d}_{\alpha}(t) = \hat{d}_{\alpha}(0) {\rm exp} \biggl( -i \frac{E_{\alpha} t}{\hbar} \biggl).
\label{TR1}
\end{eqnarray}
Thus, in the real space, the annihilation operator is given by
\begin{eqnarray}
\hat{f}_j(t) &=& \frac{1}{\sqrt{M}} \sum_{\alpha \in \Omega} \hat{d}_{\alpha}(0) {\rm exp} \biggl( i \alpha j  -i \frac{E_{\alpha} t}{\hbar} \biggl) \\
&=&  \frac{1}{\sqrt{M}} \sum_{\alpha \in \Omega} \Biggl[  \frac{1}{\sqrt{M}} \sum_{k=1}^{M} \hat{f}_k(0) {\rm e}^{-i \alpha k } \Biggl] {\rm exp} \biggl( i \alpha j  -i \frac{E_{\alpha} t}{\hbar} \biggl) \\
&=& \sum_{k=1}^{M} \hat{f}_k(0) g(j, k, t), 
\label{TR2}
\end{eqnarray} 
where we define
\begin{eqnarray}
 g(j,k,t) = \frac{1}{M} \sum_{\alpha \in \Omega}  {\rm exp} \biggl( i \alpha (j-k)  - i \frac{E_{\alpha} t}{\hbar} \biggl). 
\label{TR3}
\end{eqnarray} 
The substitution of Eq.~\eqref{TR2} into Eq.~\eqref{R1} leads to  
\begin{eqnarray}
&& \delta \hat{\rho}_j = \sum_{k=1}^{M} \sum_{l=1}^{M} \hat{f}_{k}^{\dagger}(0) \hat{f}_{l}(0) F(j,k,l,t) - \nu,  \\
&& F(j,k,l,t) =   g(j,k,t)^*  g(j,l,t).
\end{eqnarray} 
Thus, we can derive the time evolution of the surface-height operator \eqref{R2}:
\begin{eqnarray}
&& \hat{h}_{j}(t) = \sum_{k=1}^{M} \sum_{l=1}^{M} G(j,k,l,t) \hat{f}_{k}^{\dagger}(0)  \hat{f}_{l}(0) - \nu j , \label{TR4} \\
&& G(j,k,l,t) = \sum_{m=1}^{j} F(m,k,l,t).
\end{eqnarray} 
Similarly, the spatially averaged surface-height operator is given by
\begin{eqnarray}
&& \frac{1}{M}\sum_{j=1}^{M} \hat{h}_{j}(t) = \sum_{k=1}^{M} \sum_{l=1}^{M} \bar{G}(k,l,t) \hat{f}_{k}^{\dagger}(0)  \hat{f}_{l}(0) - \frac{\nu}{2}(M+1), \label{TR5} \\
&& \bar{G}(k,l,t) = \frac{1}{M}\sum_{j=1}^{M} G(j,k,l,t).
\end{eqnarray} 
In the following, we denote $\hat{f}_{q}(0)$ by $\hat{f}_{q}$ for brevity.

By using Eq.~\eqref{TR4}, we obtain the following quantum averages:
\begin{eqnarray}
\langle \hat{h}_{j}(t)^2 \rangle = \nu^2 j^2 - 2 \nu j \sum_{k=1}^{M} \sum_{l=1}^{M} G(j,k,l,t) \langle\hat{f}_{k}^{\dagger}  \hat{f}_{l} \rangle 
+ \sum_{k=1}^{M}\sum_{l=1}^{M} \sum_{p=1}^{M} \sum_{q=1}^{M}  G(j,k,l,t) G(j,p,q,t) \langle \hat{f}_{k}^{\dagger}  \hat{f}_{l} \hat{f}_{p}^{\dagger}  \hat{f}_{q} \rangle, 
\label{TR6}
\end{eqnarray} 
\begin{eqnarray}
h_{\rm av}(t)  = \sum_{k=1}^{M} \sum_{l=1}^{M} \bar{G}(k,l,t)  \langle \hat{f}_{k}^{\dagger}  \hat{f}_{l} \rangle - \frac{\nu}{2}(M+1).
\label{TR7}
\end{eqnarray}  

Finally, we calculate the initial average in Eqs.~\eqref{TR6} and \eqref{TR7}. 
The initial state considered here is a staggered state with $\nu=1/2$, which is given by
\begin{eqnarray}
| \psi(0) \rangle &=& \prod_{j=1}^{M/2} \hat{s}_{2j-1}^{+}   |0 \rangle \\ 
&=& \prod_{j=1}^{M/2} \Biggl[  {\rm exp}\biggl(  i \pi \sum_{k=1}^{2j-2} \hat{f}_{k}^{\dagger} \hat{f}_{k} \biggl) \hat{f}_{2j-1}^{\dagger} \Biggl] |0 \rangle \\
&=&  {\rm exp}\biggl( i \pi \hat{D}_{M-2} \biggl) \hat{f}^{\dagger}_{M-1} {\rm exp}\biggl( i \pi \hat{D}_{M-4} \biggl) \hat{f}^{\dagger}_{M-3} \cdots {\rm exp}\biggl( i \pi \hat{D}_{2} \biggl) \hat{f}^{\dagger}_{3}  \hat{f}^{\dagger}_{1} |0 \rangle \\
&=& \biggl[ \prod_{k=2}^{M/2} {\rm exp}\biggl( i \pi \hat{D}_{2k-2} \biggl) \biggl]  \prod_{j=1}^{M/2} \hat{f}_{2j-1}^{\dagger} |0 \rangle \\
&=& {\rm exp}\biggl( i \pi \mathcal{D} \biggl)  \prod_{j=1}^{M/2} \hat{f}_{2j-1}^{\dagger} |0 \rangle, 
\label{TR8}
\end{eqnarray}
where we define an operator $\hat{D}_j=\sum_{k=1}^j \hat{f}_{k}^{\dagger} \hat{f}_{k}$ and $\mathcal{D}$ is a real number.
The quantum averages of Eqs.~\eqref{TR6} and \eqref{TR7} are calculated as
\begin{eqnarray}
\langle\hat{f}_{p}^{\dagger}  \hat{f}_{q} \rangle =
\begin{cases}
 1 & (p=q \in \mathbb{N}_{\rm odd});\\
 0 & (\rm otherwise),
\end{cases}
\label{TR9}
\end{eqnarray}  
\begin{eqnarray}
\langle \hat{f}_{k}^{\dagger}  \hat{f}_{l} \hat{f}_{p}^{\dagger}  \hat{f}_{q} \rangle =
\begin{cases}
 1 & ( k=l=p=q, k \in \mathbb{N}_{\rm odd} );\\
 1 & ( k=l, p=q, k \neq p, k \in \mathbb{N}_{\rm odd}, p \in \mathbb{N}_{\rm odd} ); \\
 1 & ( k=q, p=l, k \neq p, k \in \mathbb{N}_{\rm odd}, p \in \mathbb{N}_{\rm even} ); \\
 0 & ({\rm otherwise})
\end{cases}
\label{TR9}
\end{eqnarray} 
with the set of odd (even) natural numbers $\mathbb{N}_{\rm odd}$ ($\mathbb{N}_{\rm even}$).
As a result, we obtain
\begin{eqnarray}
 \langle \hat{h}_{j}(t)^2 \rangle &=& \frac{1}{4} j^2 -  j \sum_{k \in \mathbb{N}_{\rm odd}} G(j,k,k,t) +  \sum_{k \in \mathbb{N}_{\rm odd}} G(j,k,k,t)^2 \nonumber \\
&+& \sum_{ \substack{ k,p \in \mathbb{N}_{\rm odd} \\ k \neq p} } G(j,k,k,t) G(j,p,p,t) + \sum_{ \substack{ k \in \mathbb{N}_{\rm odd} \\  p \in \mathbb{N}_{\rm even} }}  G(j,k,p,t) G(j,p,k,t), 
\label{TR10}
\end{eqnarray} 
\begin{eqnarray}
h_{\rm av}(t)  = \sum_{k=1}^{M} \bar{G}(k,k,t)  - \frac{1}{4}(M+1).
\label{TR11}
\end{eqnarray}  
Therefore, we can numerically calculate Eqs.~\eqref{TR10} and \eqref{TR11}, and then obtain the time evolution of $w_{q}(M,t)$ of Eq.~\eqref{S_roughness}.

\begin{figure}[t]
\begin{center}
\includegraphics[keepaspectratio, width=12cm,clip]{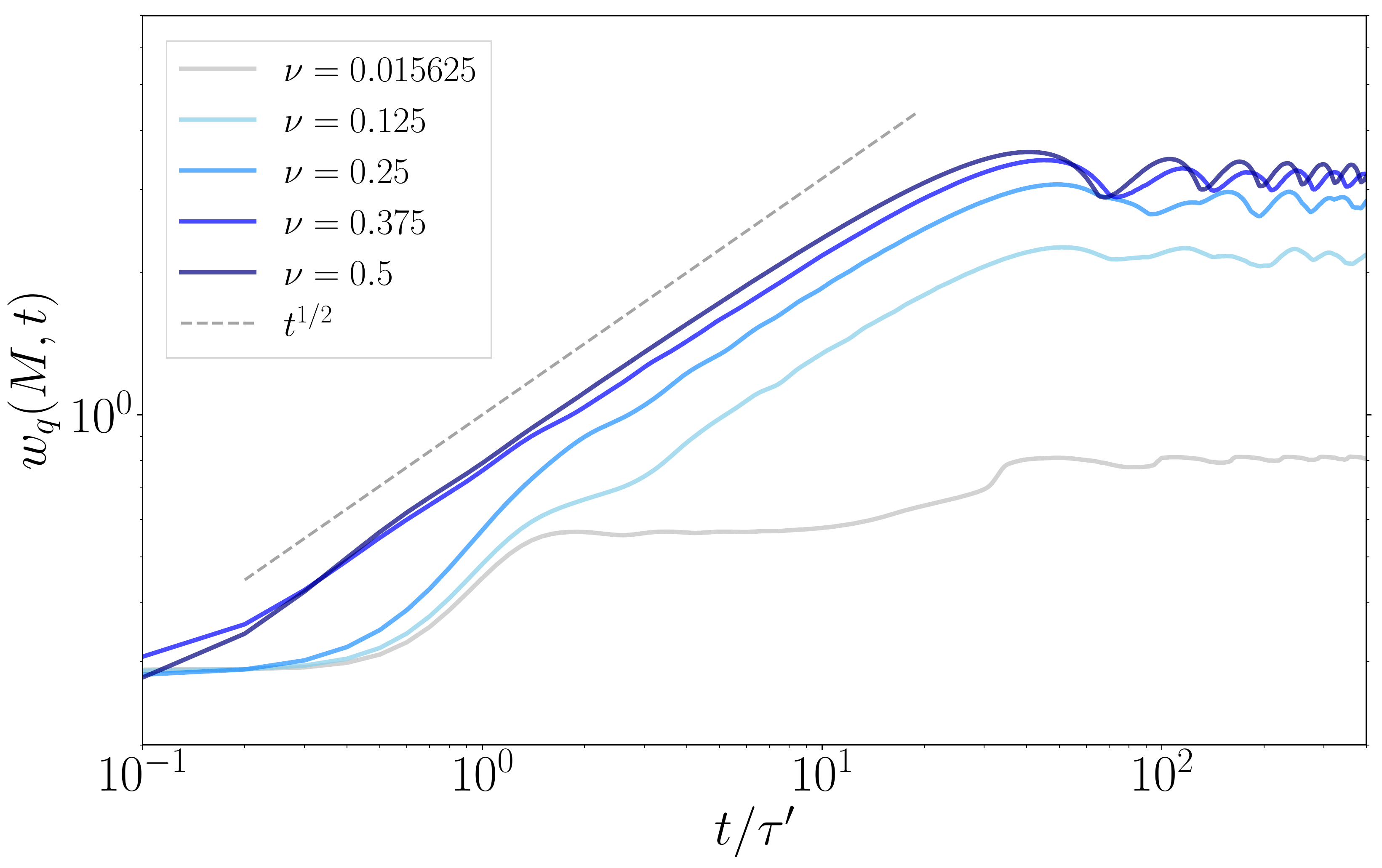}
\caption{ Time evolution of the surface roughness $w_q(M,t)$ for $\nu=0.5$, $0.375$, $0.25$, $0.125$, and $0.015625$. The system size is $M=256$, and the time is normalized by $\tau' = \hbar/2J$. At the early stage of the dynamics, the growth of the surface roughness strongly depends on the filling factor $\nu$, but at the late stage it shows the $1/2$ power-law growth except for the case with $\nu=0.015625$, which does not show any power-law growth.
\label{sup_filling} }
\end{center}
\end{figure}

\subsection{Dependence of $w_{q}(M,t)$ on the filling factor $\nu$}
Figure~4 in the main text shows the numerical result for the half filling. Here, we describe how the time evolution of $w_{q}(M,t)$ changes depending on the filling factor $\nu$. 
The initial state considered here is given by
\begin{eqnarray}
| \psi(0) \rangle &=& \prod_{j=1}^{N} \hat{s}_{[j/\nu-1]}^{+}   |0 \rangle
\end{eqnarray}
with $[\cdots]$ denoting the floor function. The particles are distributed at the equal interval for this initial state. 
By slightly modifying the calculation for $\nu = 1/2$, we can calculate time evolution of $w_{q}(M,t)$ for an arbitrary filling factor ($0<\nu<1$). 

Figure~\eqref{sup_filling} shows numerical results for $\nu=0.5$, $0.375$, $0.25$, $0.125$, and $0.015625$, from which we find that the power law growth emerging at the late stage of the dynamics is insensitive to the filling factor $\nu$ and the power exponent $\beta$ is close to $1/2$. However, the time region with the power-law growth becomes narrow as the filling factor decreases. When $\nu$ is too small, we do not find any signature of the power-law growth of the surface roughness at least for the numerically achievable system sizes.
We expect that the absence of the power law growth is attributed to too weak particle correlations.
Here, we consider the case for $\nu <1/2$ because the dynamics for $\nu > 1/2$ is equivalent to that for $1 - \nu$ due to the symmetry of the XX model.

\section{Method for extracting the exponents in the roughness growth}
From Figs.~3 and 4 in the main text, we find the clear FV scaling and obtain the power exponents  $(\alpha, \beta, z) \simeq (0.517 \pm 0.030, 0.255 \pm 0.012, 2.07 \pm 0.20)$ and $(0.500 \pm 0.003, 0.489 \pm 0.004,1.00 \pm 0.01)$ for the high- and low-filling cases, respectively. Our method to extract these values of the power exponents is based on Ref.~\cite{expoents_cal}, which is explained below. 

\subsection{Estimation of the exponents $\alpha$ and $z$}
The exponents estimated in the main text are given by:
\begin{eqnarray}
&& \alpha =  \bar{\alpha} \pm \sigma_{\alpha}, \\
&& z = \bar{z} \pm \sigma_{z},  \\
\end{eqnarray}
where $\bar{\alpha}$ and $\bar{z}$ are the averaged values, and $\sigma_{\alpha}$ and $\sigma_{z}$ are the corresponding errors. 

The averaged exponents are calculated by minimizing a deviation from the FV scaling expressed by
\begin{eqnarray}
w_{q}(M',t) = s^{-\alpha} w_{q}(sM', s^{z}t)
\end{eqnarray}
with a system size $M'$ and a constant $s$. Substituting $M'=M_{\rm ref}$ and $s=M/M_{\rm ref}$ with the reference system size $M_{\rm ref}=32$, we obtain 
\begin{eqnarray}
w_{q}(M_{\rm ref},t) = (M/M_{\rm ref})^{-\alpha} w_{q} \left(M, t (M/M_{\rm ref})^{z} \right). 
\end{eqnarray}
Then, we define the deviation $\chi(\alpha,z)$ for the FV scaling as
\begin{eqnarray}
\chi_1(\alpha,z) = \frac{1}{N_{\rm sys}} \sum_{M} \int_{t_0}^{t_1} d({\rm log}t) \frac{\displaystyle \Bigl| w_{q}(M_{\rm ref},t) - ( M / M_{\rm ref})^{-\alpha} w_{q} \left( M, t (M/M_{\rm ref})^{z} \right) \Bigl|^2}{\displaystyle w_{q}(M_{\rm ref},t)^2 }
\end{eqnarray}
with the number of system size $N_{\rm sys}$ and the fitting time span $[t_0,t_1]$. 
In the main text, we use $(t_0, t_1) = (2\tau, 400 \tau)$ and $(\tau', 16\tau')$ for the high- and low filling cases, respectively.  
By changing values of $\alpha$ and $z$, we can obtain the averaged exponents $\bar{\alpha}$ and $\bar{z}$ which minimize the deviation. 

The errors for the power exponents are calculated by the following function:
\begin{eqnarray}
P_1(\alpha,z) = \mathcal{P}_1 {\rm exp} \left( - \frac{\chi_1(\alpha,z)}{2\chi_1(\bar{\alpha},\bar{z})}  \right), 
\end{eqnarray}
where $\mathcal{P}_1$ is the normalization constant determined by $\int d\alpha dz P_1(\alpha,z) = 1$. 
Then, we define the errors $\sigma_{\alpha}$ and $\sigma_z$ by
\begin{eqnarray}
&&\sigma_{\alpha} = \int d\alpha dz P_1(\alpha,z) (\alpha - \bar{\alpha})^2, \\
&&\sigma_{z} = \int d\alpha dz P_1(\alpha,z) (z - \bar{z})^2. \\
\end{eqnarray}

\subsection{Estimation of the exponents $\beta$}
The power exponent $\beta$ is estimated by fitting the power-law function $c t^{\beta}$ with the constant $c$ to the roughness growth data for the largest system size $M_{\rm max}$. We denote the power exponent by
\begin{eqnarray}
&& \beta =  \bar{\beta} \pm \sigma_{\beta}, 
\end{eqnarray}
with the averaged value $\bar{\beta}$ and the error $\sigma_{\beta}$.

We obtain the averaged power exponent $\bar{\beta}$ by minimizing the following function:
\begin{eqnarray}
&& \chi_2 (\beta, c) = \int_{t_0}^{t_1} d({\rm log}t) \frac{\displaystyle \Bigl| w_{q}(M_{\rm max},t) - c t^{\beta}  \Bigl|^2}{\displaystyle w_{q}(M_{\rm max},t)^2 }.
\end{eqnarray}
In the main text, we use $(t_0, t_1) = (2\tau, 100 \tau)$ and $(\tau', 10\tau')$ for the high- and low filling cases, respectively.  
In this procedure, we can obtain $\bar{\beta}$ and $\bar{c}$ which minimize $\chi_2 (\beta,c)$.
Then, the error for $\beta$ is defined by
\begin{eqnarray}
\sigma_{\beta} = \int  d\beta dc P_2(\beta,c) (\beta - \bar{\beta})^2,
\end{eqnarray} 
where we use the function $P_2(\beta,c)$ given by
\begin{eqnarray}
P_2(\beta,c) = \mathcal{P}_2 {\rm exp} \left( - \frac{\chi_2(\beta, c)}{2\chi_2(\bar{\beta},\bar{c})}  \right)
\end{eqnarray}  
with the normalization constant $\mathcal{P}_2$ determined by $\int d\beta dc P_2(\beta,c) = 1$.

\end{document}